\documentclass[prb, twocolumn,superscriptaddress,showpacs]{revtex4}  
\usepackage{amssymb}
\usepackage{times}
\usepackage{amsmath}
\usepackage{bbm}
\usepackage{graphicx}
\usepackage{color}
\usepackage{enumerate}

\begin{document}

\author{Philipp Werner}
\affiliation{Department of Physics, University of Fribourg, 1700 Fribourg, Switzerland}
\author{Martin Eckstein}
\affiliation{Max Planck Research Department for Structural Dynamics, University of Hamburg-CFEL, Hamburg, Germany}

\title{Relaxation dynamics of the Kondo lattice model}

\date{\today}

\hyphenation{}

\begin{abstract}
We study the relaxation properties of the Kondo lattice model using the nonequilibrium dynamical mean field formalism in combination 
with the non-crossing approximation. The system is driven out of equilibrium either by a magnetic field pulse which perturbs the local 
singlets, or by a sudden quench of the Kondo coupling. For relaxation processes close to thermal equilibrium (after a weak perturbation), 
the relaxation time increases substantially as one crosses from the local moment regime into the heavy Fermi liquid. A strong perturbation,
which injects a large amount of energy, can rapidly 
transform the heavy Fermi liquid into a local moment state. 
%
Upon cooling, the heavy Fermi liquid reappears in  a two-stage relaxation, where the first step opens the Kondo gap and the second step corresponds to a slow approach of the equilibrium state via a nonthermal pathway. 

\end{abstract}

\pacs{71.10.Fd, 71.27.+a}

\maketitle

\section{Introduction}

Heavy Fermion compounds contain strongly interacting $f$-electrons which hybridize with extended $s$-, $p$- and $d$-electrons. 
The $f$-electrons are in a well-defined charge state, and at high temperature they 
act as local magnetic moments which scatter the conduction electrons. At low temperature, the hybridization between $f$- and 
conduction electrons 
can lead to the emergence of nontrivial electronic phases,
such as the Kondo insulating state at half filling, or a strongly renormalized Fermi liquid in the 
doped case.\cite{Kuramoto2000, Coleman2007}

A simple model which captures
essential aspects of the physics of these materials is the Kondo lattice model.\cite{Doniach1977} In this model, charge 
fluctuations of the $f$-electrons are completely suppressed. 
The localized degrees of freedom are described by spins $S=1/2$, which are coupled to the 
spin of the conduction electrons 
at the same site via an exchange interaction $J$ (the Kondo coupling). 
For antiferromagnetic coupling ($J>0$), a large value of $J$ favors the formation of singlets on each site. At half-filling, this leads to the opening of a (pseudo-)gap in the spectral function of the conduction electrons. In the ferromagnetic Kondo lattice model ($J<0$),
a metallic phase is realized at small coupling, with a phase transition (at half filling) to an insulating state at 
a critical value of $J$.\cite{Werner06Kondo} 

The most interesting behavior is found away from half-filling, 
where an antiferromagnetic interaction with the localized moments leads to a renormalized bandstructure at low temperature, 
which resembles a flat band hybridized with the wide band of the conduction electrons. This Fermi liquid is characterized by 
strong mass renormalizations and a ``large" Fermi surface, 
i.e., {\it both} conduction 
electrons and localized moments participate in the formation of the Fermi liquid state, 
and the Luttinger volume thus contains the total number of $c$- and $f$-electrons.%
\cite{Oshikawa2000}
As the temperature is raised or the coupling $J$ is reduced, a crossover occurs to a metallic state with a blurred, ``small" Fermi surface,
whose Luttinger volume contains the conduction electrons only.
This physics has been beautifully demonstrated in a recent series of papers by Otsuki and collaborators, based on the dynamical 
mean field approximation (DMFT).\cite{Otsuki09a, Otsuki09b}

In the present
paper, we use the nonequilibrium extension of DMFT to investigate the real-time dynamics of the Kondo lattice 
model under strong nonequilibrium conditions. In particular, we are interested in the timescales on which the system
can undergo a transition between states with a large and small Fermi surface. 
In practice, it is easy to perturb the 
system so strongly that  the heavy Fermi liquid is destroyed after re-thermalization at higher energy. 
We will implement such a perturbation by a sudden change of the interaction $J$, or a short magnetic field pulse, 
and investigate the crossover into the state with small Fermi surface in real time. The transition back to the heavy Fermi
liquid is then achieved upon cooling, by coupling the system to a dissipative environment which absorbs the energy 
injected into the system by the perturbation.  
Although 
the maximal cooling rate 
is limited by the coupling strength, the 
way in which the Fermi liquid state is approached and the timescale for this process  
can still reveal intrinsic properties of 
the Kondo lattice model. To understand those relaxation times, we also consider the relaxation of the system close 
to equilibrium, by perturbing the local singlets with a weak magnetic field pulse. Due to the small amount of energy 
injected by such a weak pulse, the time evolution takes place only within one given phase and allows us to extract 
the equilibrium relaxation rate in the various temperature and doping regimes, and to connect this quantity to the 
presence or absence of strongly renormalized quasi-particles.

\begin{figure}[t]
\centering
\includegraphics[width=0.7\linewidth]{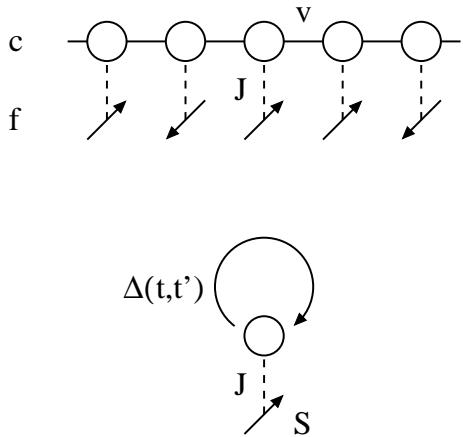}
\caption{Top panel: illustration of the Kondo lattice model, describing conduction band electrons $c$ hopping with matrix element $v$ between orbitals (circles) and interacting via an exchange coupling $J$ with the spin of the localized $f$-electrons (arrows). Bottom panel: dynamical mean field approximation of the Kondo lattice model, consisting of one $c$-electron orbital and the associated $f$-electron spin. The $c$-electrons couple to a self-consistently determined bath of noninteracting 
sites, with hybridization function $\Delta(t,t')$.}
\label{kondo_fig}
\end{figure}

\section{Model and method}
\label{methodsection}

The spin-$\frac{1}{2}$ Kondo lattice model describes conduction electrons $c$ interacting with localized electrons $f$. If the $f$-orbitals are half-filled and fluctuations into empty or doubly occupied states are energetically very expensive, only the spin degree of freedom $\mathbf{S}=\frac{1}{2}\psi_f^\dagger\boldsymbol{\sigma}\psi_f$ remains in a low-energy model, while the charge fluctuations are suppressed [$\psi_f^\dagger=(f^\dagger_\uparrow, f^\dagger_\downarrow)$]. The Hamiltonian of the Kondo lattice model then becomes
\begin{equation}
\label{kondo}
H=-\sum_{i\ne j,\sigma} v_{ij} c^\dagger_{i,\sigma}c_{j,\sigma}+\mu\sum_{i,\sigma} n_{i,\sigma}+J\sum_i 
\boldsymbol{S}_i
\cdot 
\boldsymbol{s}_i.
\end{equation}
Here, the first term corresponds to the kinetic energy of the conduction electrons, the second term gives the chemical potential 
contribution of the conduction electrons ($n_{i,\sigma}=c^\dagger_{i,\sigma} c_{i,\sigma}$), and the last term describes the interaction 
of the spins $\boldsymbol{s}_i= \tfrac12\psi_{c,i}^\dagger \boldsymbol{\sigma} \psi_{c,i}$
of the conduction electrons with the localized electrons via the Kondo coupling $J$  [$\psi_c^\dagger=(c^\dagger_\uparrow, c^\dagger_\downarrow)$]. 
In this study, we will restrict our attention 
to antiferromagnetic $J$ and to paramagnetic solutions.

To investigate the properties of this model, we use dynamical mean field theory (DMFT).\cite{Georges96} This approximate method, which becomes exact in the limit of infinite coordination 
number,\cite{Metzner89} maps the lattice model onto a self-consistent solution of a quantum impurity model. 
The mapping can be applied to nonequilibrium situations\cite{Schmidt2002, Freericks2006} 
by reformulating the theory on the Keldysh time contour $\mathcal{C}$.
As illustrated in Fig.~\ref{kondo_fig}, the impurity consists of a $c$-electron orbital coupled to a spin $\boldsymbol{S}$ and a bath of non-interacting sites. The relevant properties of the bath are 
encoded in the hybridization function $\Delta(t,t')$, which describes the probability for transitions of $c$-electrons from the impurity site into the bath and back. Hence, the impurity action reads
\begin{equation}
\label{action}
\mathcal{S}
=
-i \int_\mathcal{C}\!\! dt \,H_\text{loc}(t) -i  \int_\mathcal{C} \!\!dtdt'\,\sum_\sigma
 c_\sigma^\dagger(t) \Delta(t,t') c_\sigma(t') ,
 \end{equation}
where 
\begin{equation}
H_\text{loc}  = \mu(n_{\uparrow}+n_{\downarrow})
+J\boldsymbol{S} \cdot \frac{1}{2}\psi_{c}^\dagger \boldsymbol{\sigma} \psi_{c}
\label{hloc}
\end{equation}
is the local part of the Hamiltonian (\ref{kondo}),
consisting of the conduction electron orbital and the spin $\boldsymbol{S}$. 

We will consider a lattice whose noninteracting density of states is semi-elliptical with bandwidth $4v$. In this case, 
the DMFT self-consistency becomes
\begin{equation}
\Delta(t, t')=v^2 G_c(t,t'),
\label{self-consistency}
\end{equation}
where $G_c(t,t')$ denotes the $c$-electron Green function  (see Appendix).
We set $v=1$ as the unit of energy, and measure time in units of $v^{-1}$. 
The Green function $G_c(t,t')=-i\langle T_\mathcal{C} c(t)c^\dagger(t')\rangle$ must be computed numerically from the 
contour-ordered average $\langle...\rangle= \text{Tr} [T_\mathcal{C} e^{\mathcal{S} }...]/ \text{Tr} [T_\mathcal{C} e^{\mathcal{S} }]$,
using the action defined in Eq.~(\ref{action}) 
($T_\mathcal{C}$ is the time-ordering operator on the Keldysh contour $\mathcal{C}$).
 
For equilibrium calculations, several techniques are available to solve the impurity problem, among them the numerical 
renormalization group\cite{Bulla2008} and continuous-time Monte Carlo methods.\cite{solver_review} The latter come in two variants, the weak-coupling CT-J solver\cite{Otsuki07} and the 
hybridization expansion approach (CT-HYB).\cite{Werner06Kondo}
In a CT-J simulation, the partition function of the impurity model is expanded in powers of $J$, so that the Monte 
Carlo configurations consist of arbitrary sequences of spin-flip processes, with weight proportional to a determinant of a matrix of bath Green functions. In a CT-HYB 
calculation, the local problem $H_\text{loc}$ is solved exactly, while the expansion of the partition function is 
done in powers of the hybridization function $\Delta$. Here, the weight is proportional to the determinant of a matrix of hybridization functions. 

Neither CT-J nor CT-HYB suffers from a sign problem in equilibrium calculations. Monte Carlo simulations 
on the real-time Keldysh contour, however, 
lead to a dynamical 
sign problem,\cite{Werner09, Werner10} which restricts the simulations to rather short times.  
Since we are interested in both the transient dynamics of the Kondo lattice model {\it and} the long-time relaxation towards an equilibrium state, we use the non-crossing 
approximation (NCA) as an impurity solver. Similar to CT-HYB, NCA is based on an expansion of the impurity partition function in powers of the hybridization 
functions. But rather than combining various (crossing and non-crossing) diagrams into a determinant, only the non-crossing diagrams are retained and summed up 
analytically via a Dyson equation. The implementation of the NCA impurity solver on the Keldysh contour has been explained in Ref.~\onlinecite{Eckstein10nca}, and 
we use the same procedure with the local Hamiltonian $H_\text{loc}$ represented  as a $8\times 8$ block matrix in the basis $|S; n_{\uparrow}; n_{\downarrow}\rangle$, 
with $S=\uparrow, \downarrow$ and $n_{\sigma}=0,1$. 
Other formulations of (extendend) NCA for the Kondo lattice model have been previously
used.\cite{Pruschke1995} An advantage of the present formulation with an $8$-dimensional local problem is that it 
captures the formation of local singlets at the $0$th order of the approximation.\cite{footnote00}
The NCA solver does not suffer from a sign problem, so that the computational effort grows polynomially (like the 
third power) with the maximum simulation time. We converge the DMFT equations on the real-time contour time-step by time-step, using the procedure detailed in 
Ref.~\onlinecite{Eckstein10quench}.

In order to characterize the various equilibrium and nonequilibrium phases, we compute
both static observables, such as the momentum distribution, and the frequency dependent
spectral function. For a general time-evolving state, the latter can be defined as 
\begin{equation}
A(\omega,t)=
-\frac{1}{\pi} \text{Im}
\int_{t}^{\infty} dt' e^{i\omega (t'-t)}G^\text{ret}(t',t)
\label{A_omega_t},
\end{equation}
where $G^\text{ret}(t,t')=-i\Theta(t-t')\langle\{c(t),c(t')\}\rangle$ is the retarded Green function.
In practice, the time-integral is limited to some maximal time $t'=t_\text{max}$, which could 
lead to artificial oscillations in the results. Below, this effect is well controlled with a 
suitably large cutoff $t_\text{max} \approx 50 $.

In principle, there is always some arbitrariness in the definition of a nonequilibrium spectral function,
and in general, the Fourier transform Eq.~(\ref{A_omega_t}) need not even be positive.
However, in particular when the inverse width of the spectral features is small compared 
to the timescale on which $A(\omega,t)$ changes, the function (\ref{A_omega_t}) is closely 
related to a 
time-resolved photoemission and inverse photoemission spectrum.\cite{TRPES} 
Moreover, $A(\omega,t)$ constitutes a complete representation  of the local Green function,
and for an equilibrium state $A(\omega) \equiv A(\omega,t)$ becomes time-independent and 
reduces to the conventional definition.

\begin{figure}[t]
\centering
\includegraphics[angle=-90, width=0.9\linewidth]{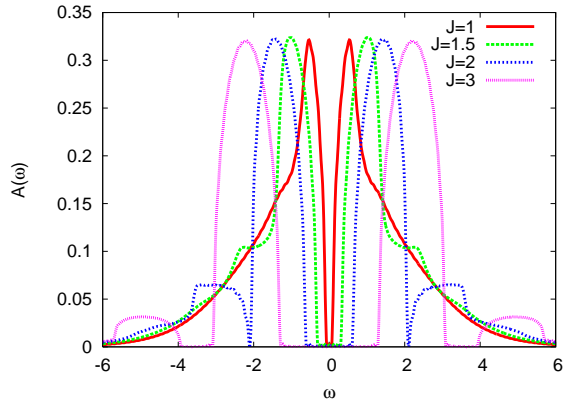}
\caption{
Spectral functions of the $n_c=1$ Kondo insulator at $\beta=50$ and indicated values of $J$.
}
\label{fig_kondo_ins}
\end{figure}

\begin{figure}[t]
\centering
\includegraphics[angle=-90, width=0.9\linewidth]{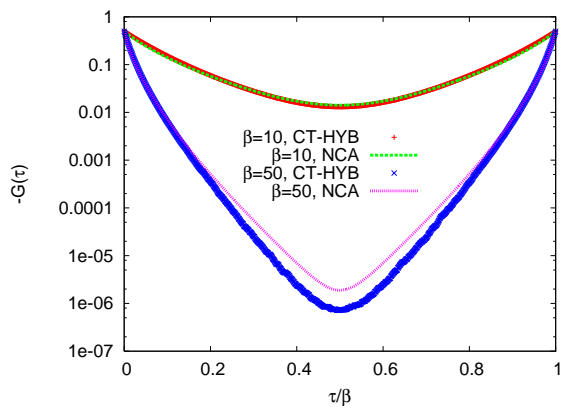}
\caption{
Comparison between 
Green functions from CT-HYB and NCA 
for $J=1.5$, $n_c=1$, and indicated temperatures.
}
\label{fig_test_nca}
\end{figure}

\section{Results}
\label{resultssection}
\subsection{Equilibrium properties}

In order to set the stage for the study of the relaxation dynamics and to test the quality of the NCA approximation, we first compute 
various results for the Kondo lattice model in equilibrium. Figure~\ref{fig_kondo_ins} shows the conduction electron spectral function 
(\ref{A_omega_t})
at half-filling ($\mu=0$, $n_c=1$), for several values of $J$ and inverse temperature $\beta=50$. 
For sufficiently low temperature, the singlet formation between the conduction-electron and localized spins leads to the opening of a gap (Kondo 
insulator). The separation between the peaks is given by $E(n=2)+E(n=0)-2E(n=1)=1.5J$, where $E(n=0)=0$, $E(n=1)=-\frac{3}{4}J-\mu$ 
and $E(n=2)=-2\mu$ are the lowest energy states for the local problem ($H_\text{loc}$) with $n$ $c$-electrons.\cite{Werner06Kondo} 
The side-peaks which are split off by $J$ correspond to the insertion or removal of an electron with 
additional singlet-triplet excitations. 
Due to the exponential decay of the hybridization function in the Kondo insulating phase, we expect the NCA approximation to be rather 
accurate in this regime.\cite{Gull2010, Eckstein10nca} 
As shown in Fig.~\ref{fig_test_nca}, already for $J=1.5$ the NCA solution 
indeed provides 
a good approximation of the exact 
Green function, although
NCA slightly underestimates the size of the Kondo gap, in contrast to the Mott gap in the Hubbard model.\cite{Eckstein10nca} 

\begin{figure}[t]
\centering
\includegraphics[angle=-90, width=0.9\linewidth]{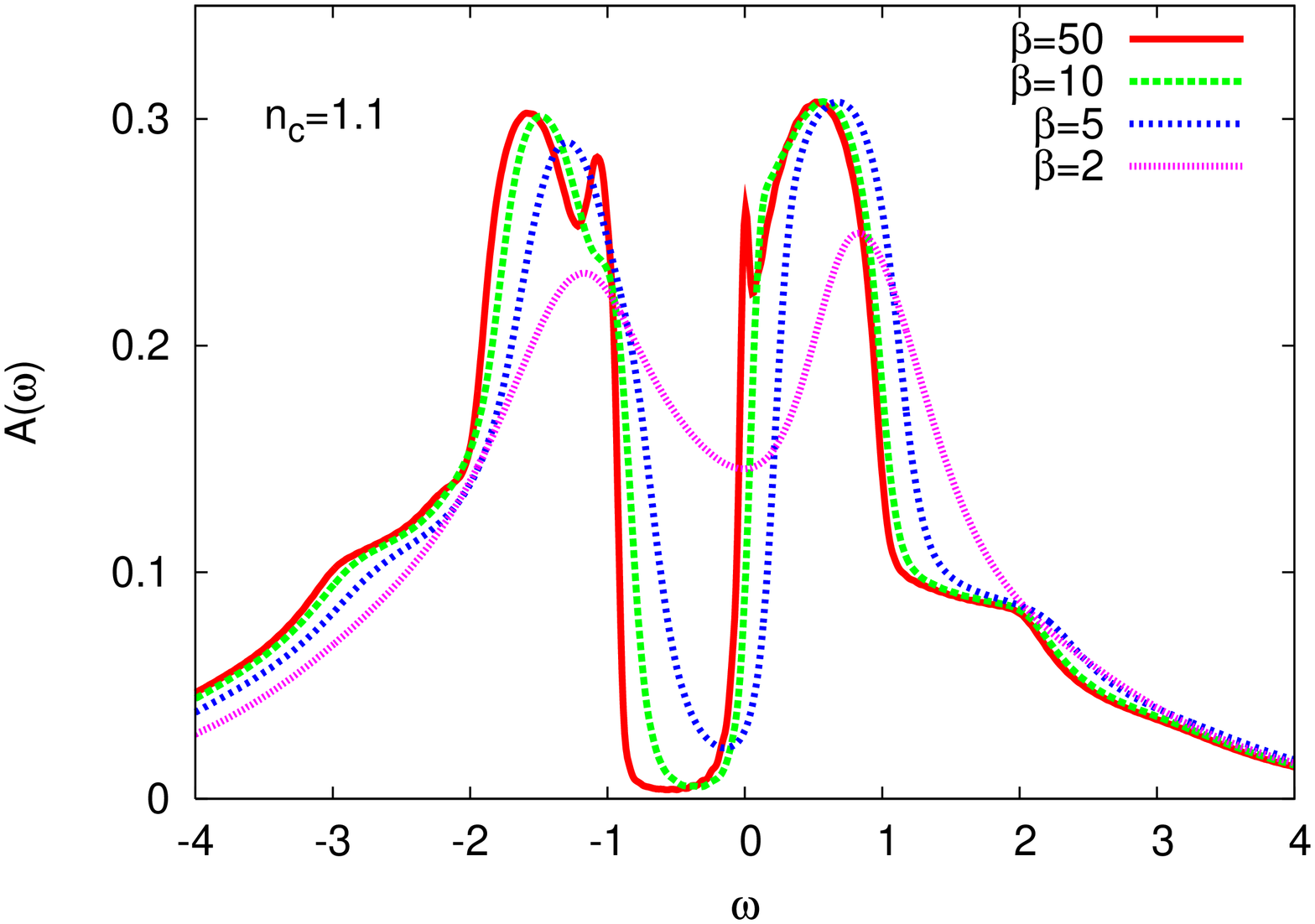}
\includegraphics[angle=-90, width=0.9\linewidth]{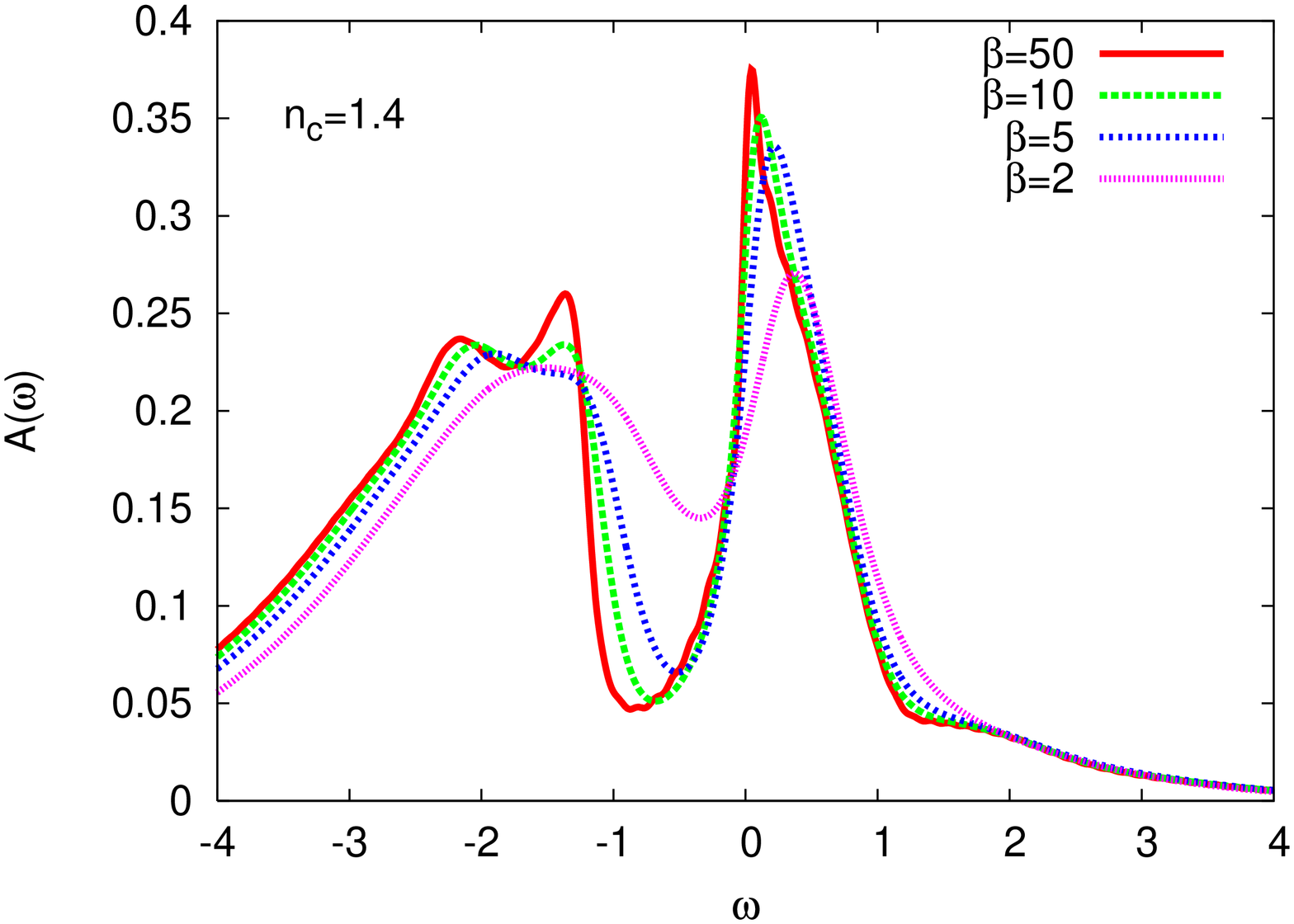}
\caption{$J=1.5$. 
Temperature-dependence 
of the spectral function for $n_c=1.1$ (top panel) and $n_c=1.4$ (bottom) panel.
}
\label{fig_kondo_doped}
\end{figure}

As one dopes the system, a narrow quasi-particle peak appears near the Fermi level at low 
temperatures. 
Figure~\ref{fig_kondo_doped} shows spectral 
functions for $J=1.5$, $n_c=1.1$ and $n_c=1.4$, and different inverse temperatures $\beta$. As the temperature is 
increased, the narrow feature disappears  ($\beta=10$), 
but the Fermi level remains at the upper edge of the 
partially filled-in Kondo 
insulator gap. Eventually, the upper gap edge moves 
away from the Fermi level ($\beta=5$) and at even higher temperatures ($\beta=2$), the gap starts to fill in. 
At larger doping, the pseudo-gap is less pronounced and the low-temperature quasi-particle peak is merged 
with the upper band. 

\begin{figure}[t]
\centering
\includegraphics[angle=-90, width=0.9\linewidth]{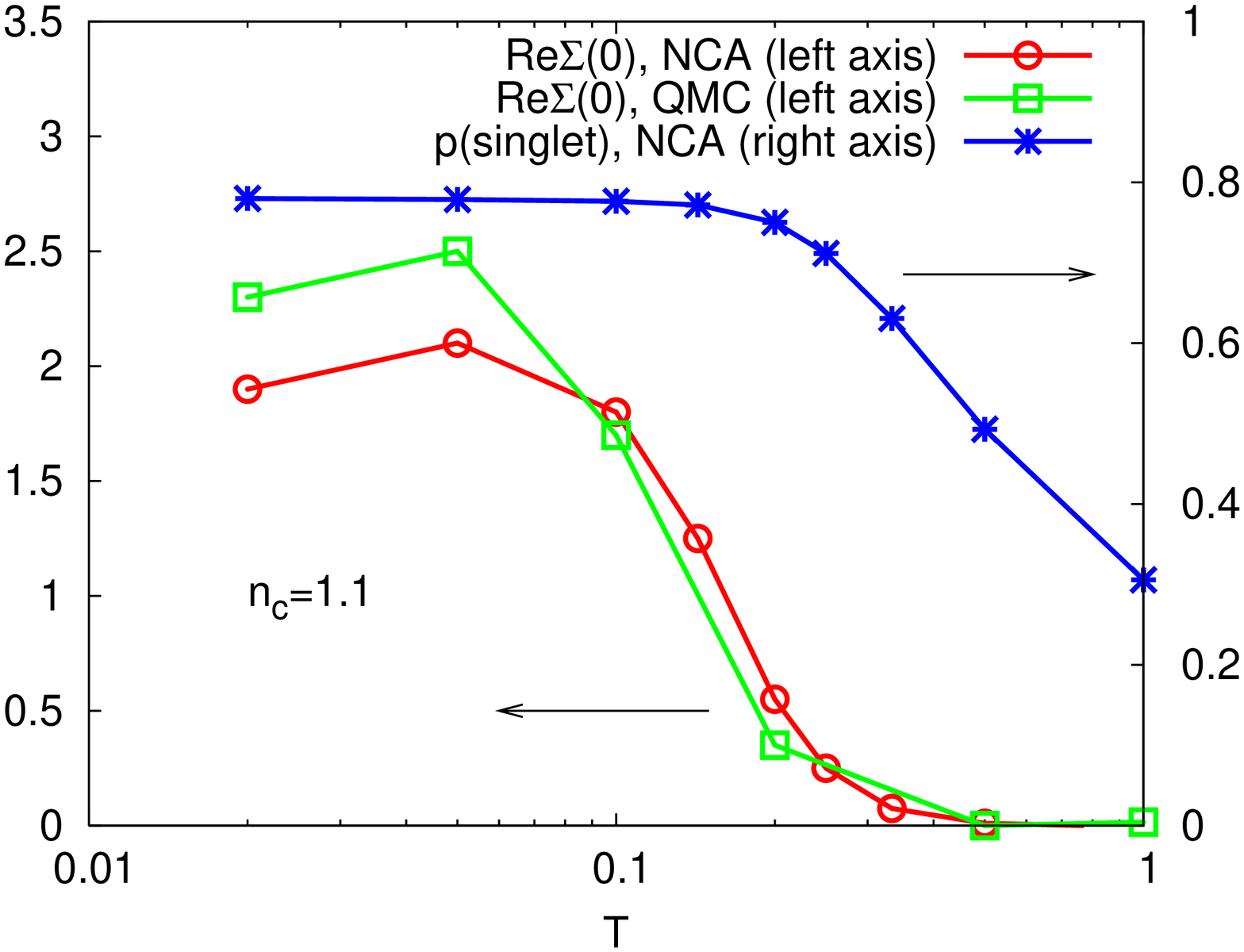}
\includegraphics[angle=-90, width=0.9\linewidth]{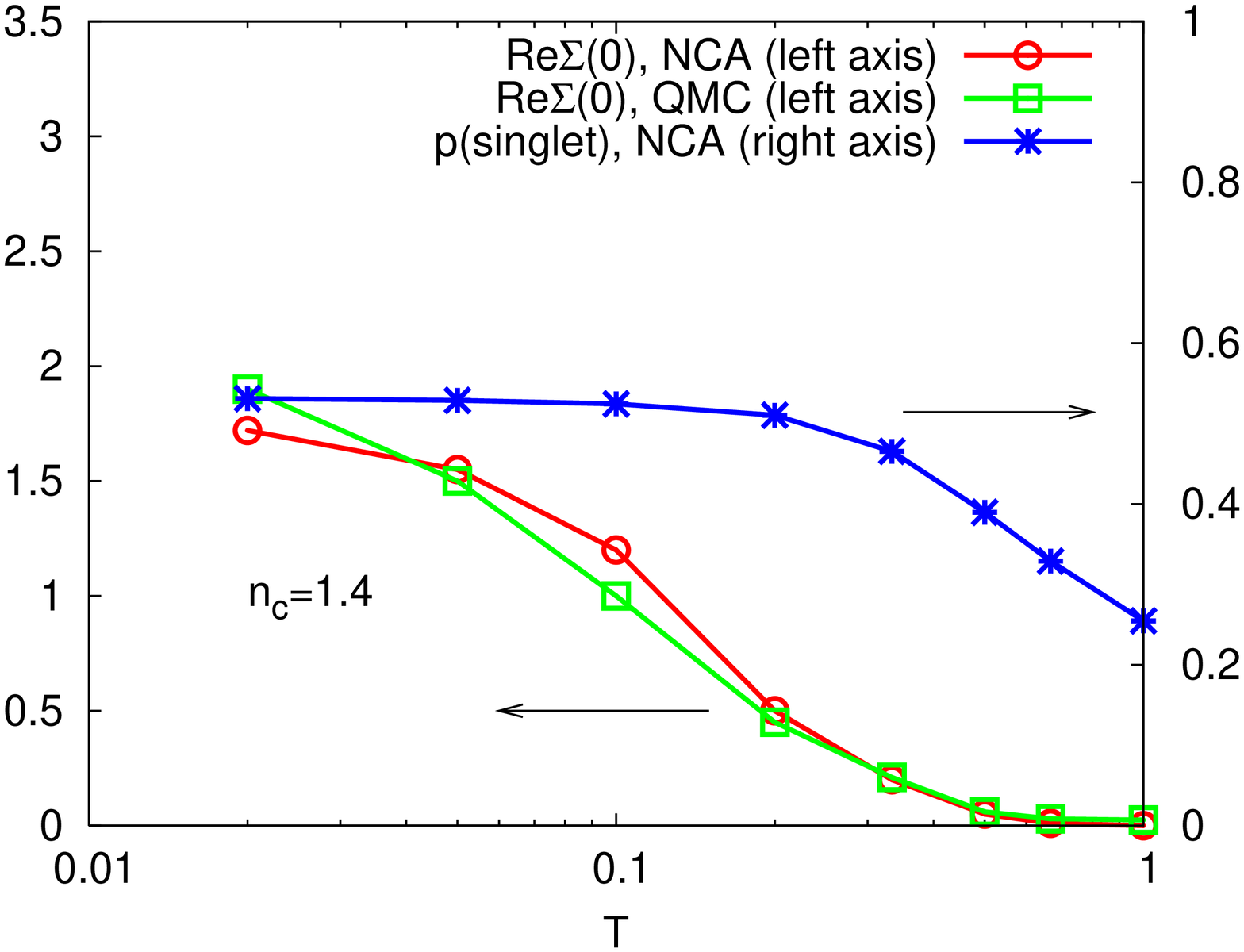}
\caption{ 
$J=1.5$. Temperature dependence of the extrapolated 
value $\text{Re}\Sigma(0)$ (left axis), showing the crossover from the 
Kondo insulator
regime to the Fermi liquid regime below $T^*$. Also plotted is the temperature dependence of the occupation of the singlet state 
(right axis), which illustrates the crossover from the local moment regime at high $T$ to the  
Kondo insulator
regime below $T_K$. The top panel shows data for $n_c=1.1$ and the bottom panel for $n_c=1.4$. 
}
\label{fig_re_sigma}
\end{figure}

To get a better understanding of the crossover between the various phases, 
we plot in Fig.~\ref{fig_re_sigma} the temperature-dependence of 
the real part of the conduction electron self-energy at frequency $\omega=0$,
$\text{Re}\Sigma(0)$, and the occupation 
$p_\text{singlet}  $ 
of the impurity singlet state 
($p_\text{singlet} = \langle P_1 (\tfrac14- \boldsymbol{s}\cdot \boldsymbol{S} )\rangle $, 
where $P_1=n_\uparrow+n_\downarrow- 2 n_\uparrow n_\downarrow$ is the projector 
on the one-particle sector of the local Hilbert space).
The behavior of $\text{Re}\Sigma(0)$ 
was used in Ref.~\onlinecite{Otsuki09a} to define the crossover scale $T^*$ 
below which a coherent Fermi liquid state is formed. The evolution of $\text{Re}\Sigma(0)$ in our case  looks 
similar to what was found in Ref.~\onlinecite{Otsuki09a} 
for a hypercubic lattice, 
and the temperature dependence is also qualitatively consistent with the numerically exact CT-HYB results for the semi-circular DOS. 
We have used
a linear extrapolation procedure to estimate $\text{Re}\Sigma(0)$
from the values $\Sigma(i\omega_n)$ at non-zero Matsubara frequencies, 
rather than the quadratic fit employed in Ref.~\onlinecite{Otsuki09a}, 
because this seems more appropriate at the relatively high temperatures considered in this study.  At the filling $n_c=1.1$, the heavy Fermi 
liquid appears below $T^*\approx 0.1$, so that we can associate the shift of the Fermi level 
in Fig.~\ref{fig_kondo_doped}
from a position within the (pseudo-)gap into the 
upper band  and the formation of a narrow resonance with the appearance of heavy quasi-particles and the formation of the large Fermi surface. 
Comparison with the 
temperature dependence
of $p_\text{singlet}$  shows that the enhanced singlet formation sets in already at a 
higher temperature, below a crossover scale $T_K\approx 0.5$. 
Hence there is a temperature range $T^*<T<T_K$ between the Fermi liquid and local moment regime, 
where the singlet formation leads to a pronounced pseudo-gap in the spectral function and the Fermi level lies inside this gap. Above $T_K$, 
$p_\text{singlet}$ decreases and the pseudogap in the spectral function starts to fill in. 
The existence of the two temperature scales $T^*$ and $T_K$ 
in the single-site DMFT solution of the Kondo lattice model has been previously discussed in Ref.~\onlinecite{Burdin00}. The crossover
temperatures are correctly reproduced by the NCA approximation and, at least in the range considered here, they show little dependence on 
doping.  

The most direct evidence for a large Fermi surface is obtained from the momentum distribution $n(\epsilon)$,
which is given by the expectation value $\langle c_{\boldsymbol{k}}^\dagger c_{\boldsymbol{k}}\rangle $ for
band energy $\epsilon=\epsilon_{\boldsymbol{k}}$
(see Appendix).
It is shown in Fig.~\ref{neps_eq} for different couplings 
$J$ at $\beta=50$ and $n_c=1.1$, and $1.4$.
At small values of $J$, a smeared-out step is visible around the location of the Fermi surface of a free conduction 
electron gas. 
The formation of the heavy quasi-particle band leads to a step in the distribution function $n(\epsilon)$
at a different location and with a smaller size, corresponding to a shift of the Fermi surface
and  a reduction of the quasiparticle weight. 

We conclude from these results that our NCA approximation, which exactly treats an 8-dimensional local problem, provides a qualitatively correct description 
of the Kondo insulator, the heavy Fermi liquid and local moment regimes in the doped model, and of the various crossover phenomena.  In the following, we 
will use this approach to study the relaxation properties of the Kondo lattice model in the different parameter regimes.

\begin{figure}[t]
\centering
\includegraphics[angle=-90, width=0.9\linewidth]{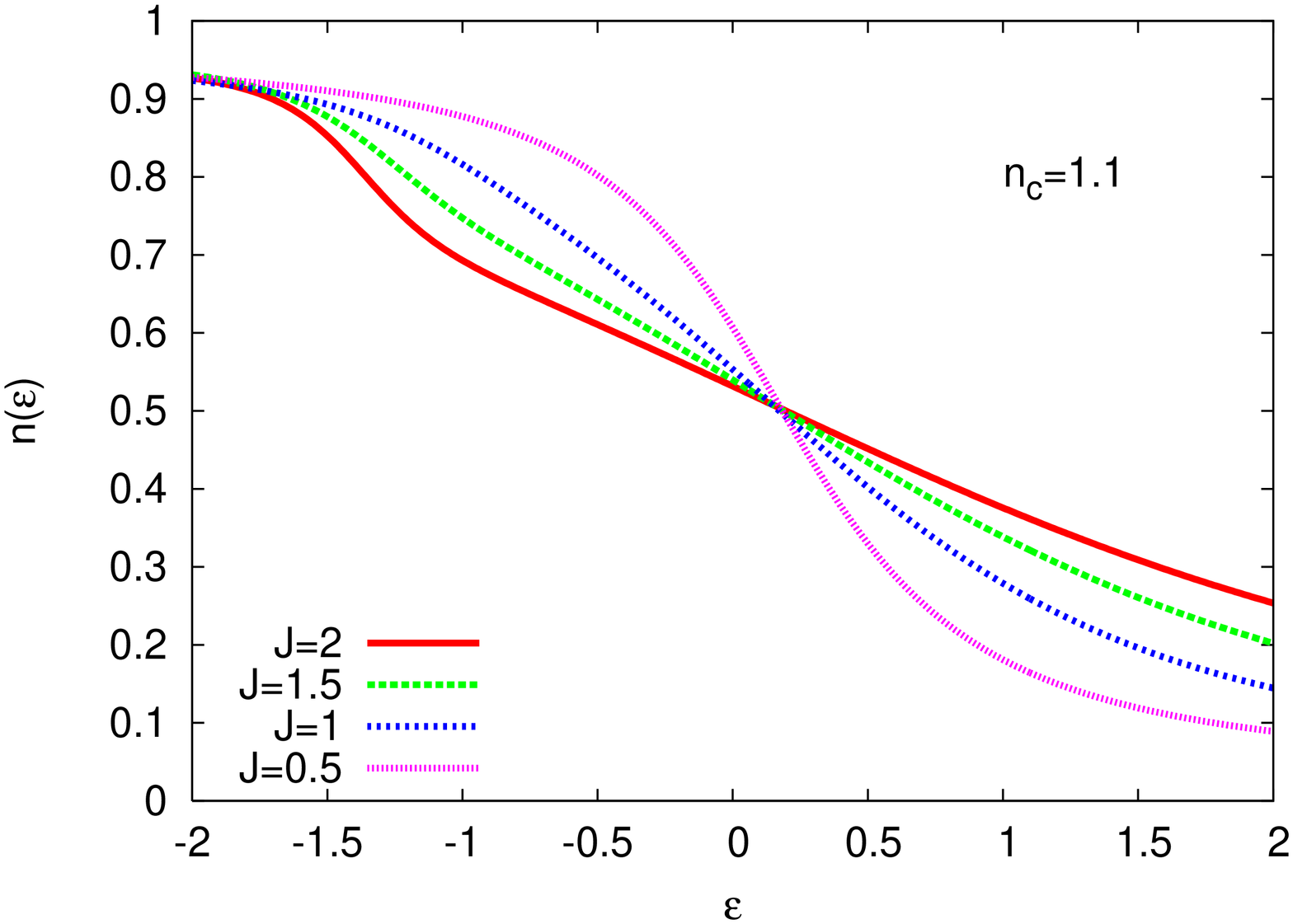}
\includegraphics[angle=-90, width=0.9\linewidth]{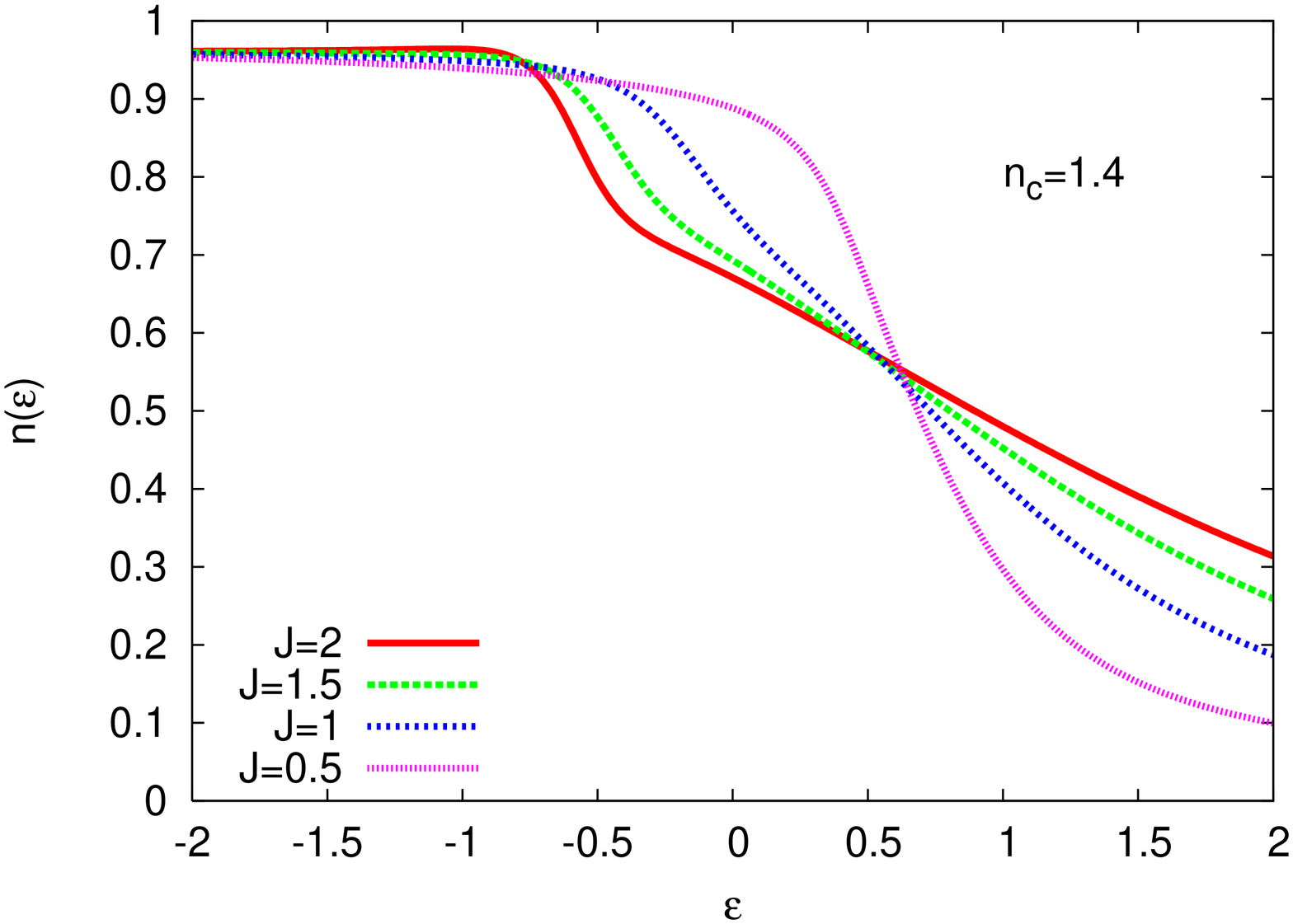}
\caption{Distribution function $n(\epsilon)$ for the antiferromagnetic Kondo lattice model at $\beta=50$. 
The top panel corresponds to a filling $n_c=1.1$ and the bottom panel to $n_c=1.4$. For large $J$, one observes the formation of a small kink in $n(\epsilon)$, corresponding to the 
large
 Fermi surface, where the heavy quasiparticle band crosses the Fermi energy.}
\label{neps_eq}
\end{figure}

\subsection{Relaxation close to thermal equilibrium}
\label{secb}

We next compute the relaxation times of the system after it is weakly perturbed in one of 
the various parameter regimes discussed above. This will later on be important  to 
understand the behavior of the system  after a strong perturbation, for times long enough that a new equilibrium state is approached. 
The relaxation times are still entirely determined by 
equilibrium properties of the corresponding electronic phase, which is not left by the weak perturbation.
Technically, the following investigation could thus be done by computing a suitable linear response quantity 
within the Matsubara formalism, but it is more convenient to do an explicit real-time calculation,
which avoids analytical continuations and also does not require the evaluation of vertex functions of the impurity model.

Specifically, we consider the relaxation of the system after a short and weak magnetic field pulse, 
which acts only on the localized spin by means of an additional perturbation $h_x(t)S_x$ in the Hamiltonian (\ref{hloc}). 
The pulse is of the form 
\begin{equation}
h_x(t)=h\sin^2(\pi t/t_\text{pulse})  
\label{hpulse}
\end{equation}
for $t<t_\text{pulse}$ and $h_x(t)=0$ for $t>t_\text{pulse}$. We use pulses of duration $t_\text{pulse}=1.5$ and small field strength $h$, such that 
the heating effect is comparatively weak
and we can study the relaxation time within the three temperature regimes $T<T^*$ (``Fermi liquid", FL), $T^*<T<T_K$ (``Kondo insulator", KI) 
and $T>T_K$ (``local moment", LM). We will focus on the time evolution of $p_\text{singlet}$. Other observables, such as the double-occupancy on the $c$-site, 
or the distribution function $n(\epsilon, t)$ seem to give consistent results for the relaxation time.

Because the pulse is acting only on one of the two spins which form the singlet (the localized spin $\boldsymbol{S}$), 
it perturbs the singlet and leads to an initial decrease in $p_\text{singlet}$. 
This decrease 
is followed by a complicated transient evolution up 
to about $t\approx 10$, after which the system eventually settles into an exponential relaxation 
towards a new thermal equilibrium state at somewhat higher temperature. 
To measure the relaxation time $\tau$, we fit this long-time behavior with an exponential function
$p_\text{singlet}(t) = p_\text{singlet}(t=\infty) + A\exp(-t/\tau)$.
For several sets of parameters
we have cross-checked that within  the accuracy of our calculation the extrapolated final value $p_\text{singlet}(t=\infty)$ 
obtained from the fit corresponds to the value computed independently by assuming thermalization at constant energy.

\begin{figure}[t]
\centering
\includegraphics[angle=-90, width=0.9\linewidth]{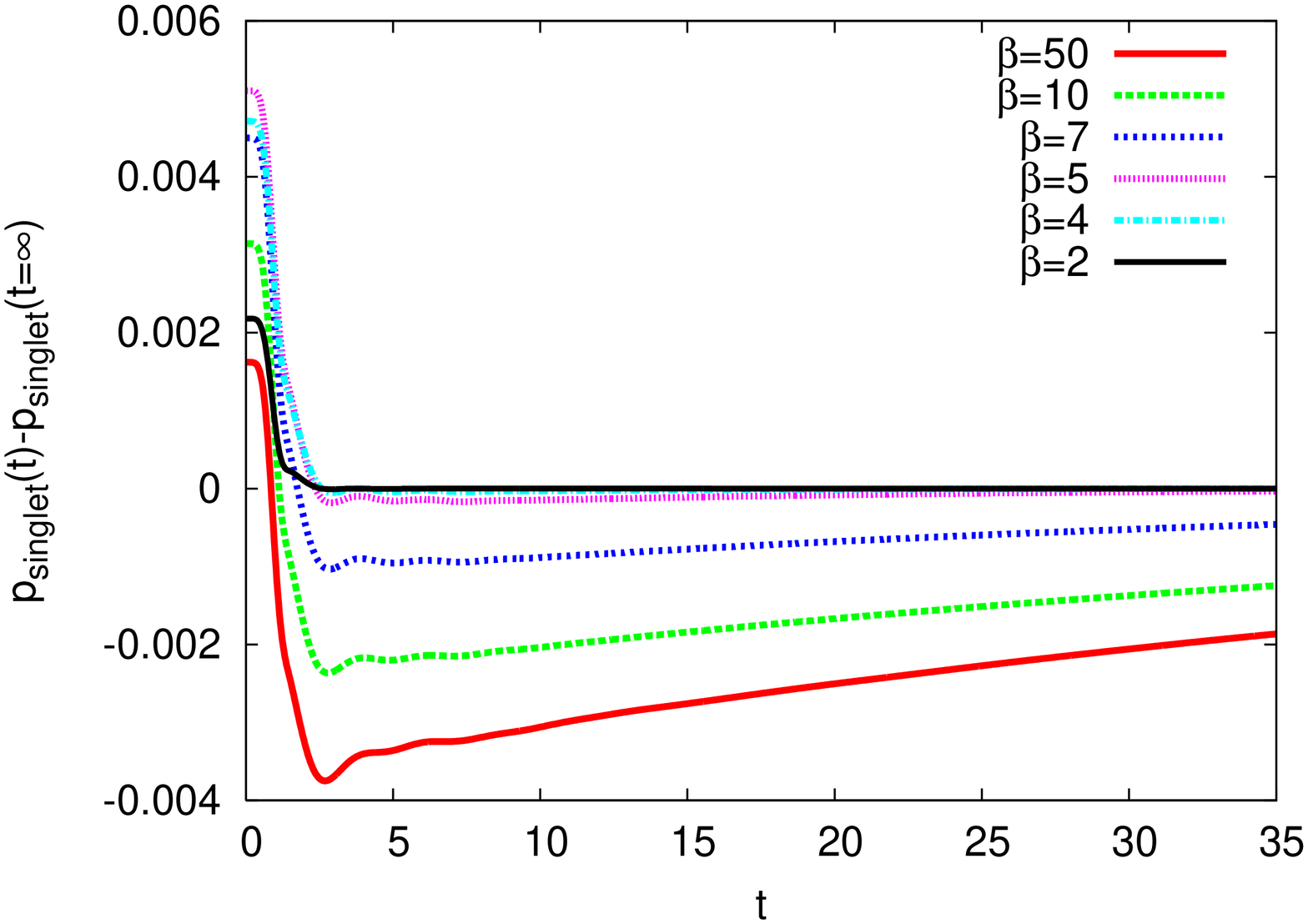}
\includegraphics[angle=-90, width=0.9\linewidth]{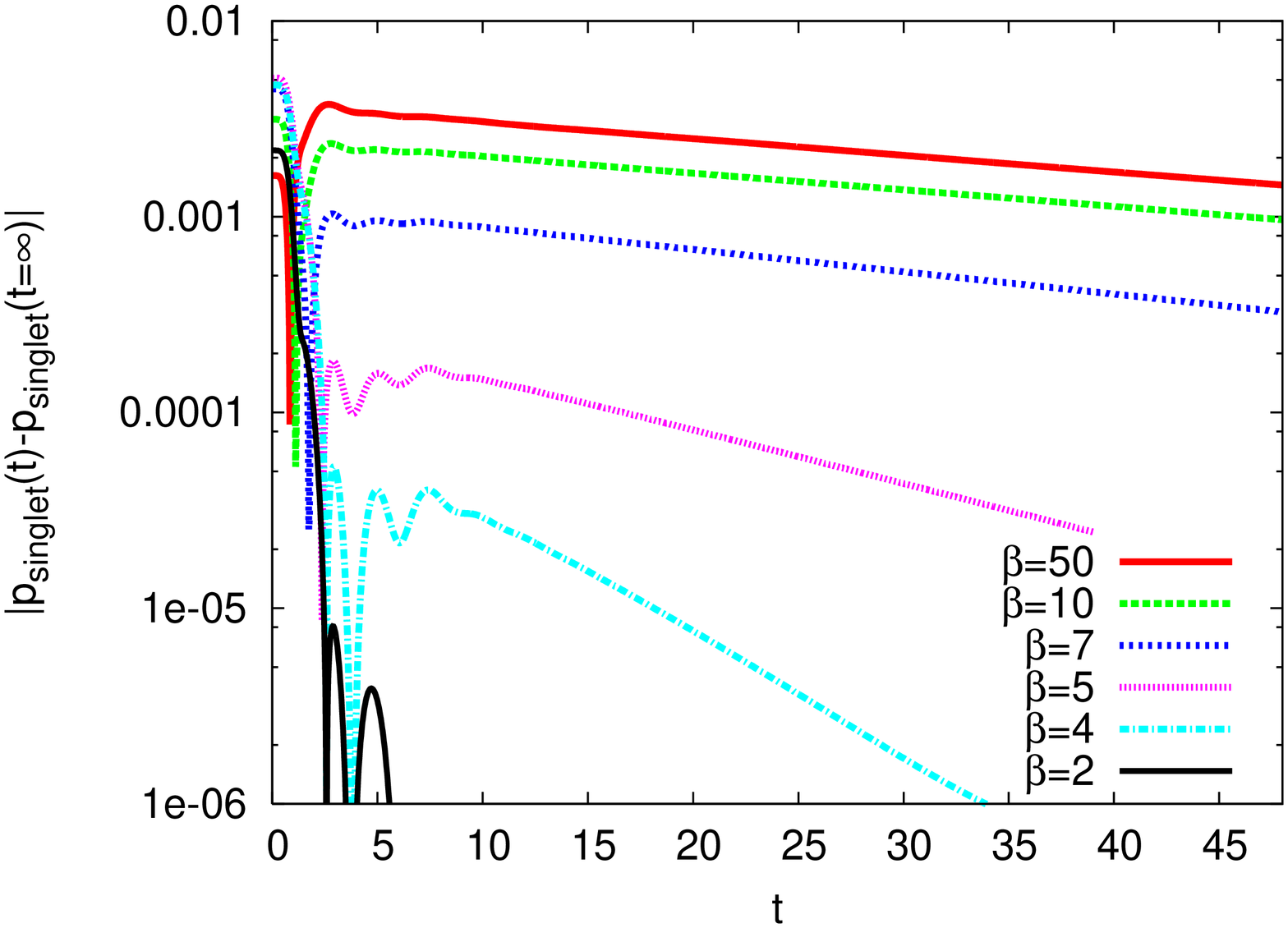}
\caption{Relaxation of $p_\text{singlet}$ after a $h$-field pulse of strength $h=0.1$, as a function of temperature. The initial state is the equilibrium state for 
indicated temperatures, $n_c=1.1$, $J=1.5$. 
After the pulse, the system with $\beta=50$ relaxes to a thermal state with inverse temperature $\approx 15$, and the system 
with $\beta=5$ to a thermal state with inverse temperature $\approx 4.8$.}
\label{fig_relax_T}
\end{figure} 

Figure~\ref{fig_relax_T} shows the time-evolution of $p_\text{singlet}(t)-p_\text{singlet}(t=\infty)$ after a pulse of strength $h=0.1$ and duration $t_\text{pulse}=1.5$, for $J=1.5$ and $n_c=1.1$. The initial state is the equilibrium state at different temperatures. 
In the FL regime ($T\lesssim 0.1$), the transient is relatively smooth, and the exponential relaxation towards the equilibrium state
becomes remarkably slow. In the KI regime $1/7 \lesssim T \lesssim 1/3$, three things happen: (i)~the transient exhibits a 
plateau
with oscillations whose period is roughly  proportional to $1/J$,
(ii) the exponential relaxation becomes substantially faster with increasing temperature, 
and (iii) the exponential relaxation sets in from a value of $p_\text{singlet}$ which is much closer to the thermal value 
than in the FL regime. Finally, in the LM regime, the relaxation is so fast that after the initial transient (which seems to 
take about $t\approx 10$, independent of temperature), the system has already thermalized. Within the accuracy of 
our simulation, 
it then no longer makes sense 
to  fit an exponential to the long-time evolution.

\begin{figure}[t]
\centering
\includegraphics[angle=-90, width=0.9\linewidth]{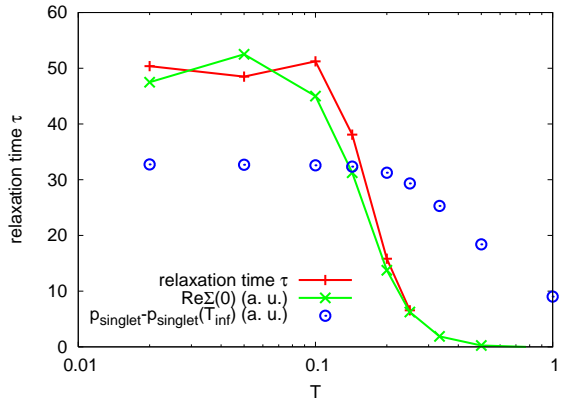}
\caption{Relaxation time $\tau$ after a $h$-field pulse of strength $0.1$, as a function of temperature. The initial state is the equilibrium state for the indicated 
temperatures 
and $n_c=1.1$, $J=1.5$. 
For comparison, we also plot $\text{Re}\Sigma(0)$ and $p_\text{singlet}$ (with arbitrary rescaling).
}
\label{fig_relax_vs_resigma}
\end{figure} 

\begin{figure}[t]
\centering
\includegraphics[angle=-90, width=0.49\linewidth]{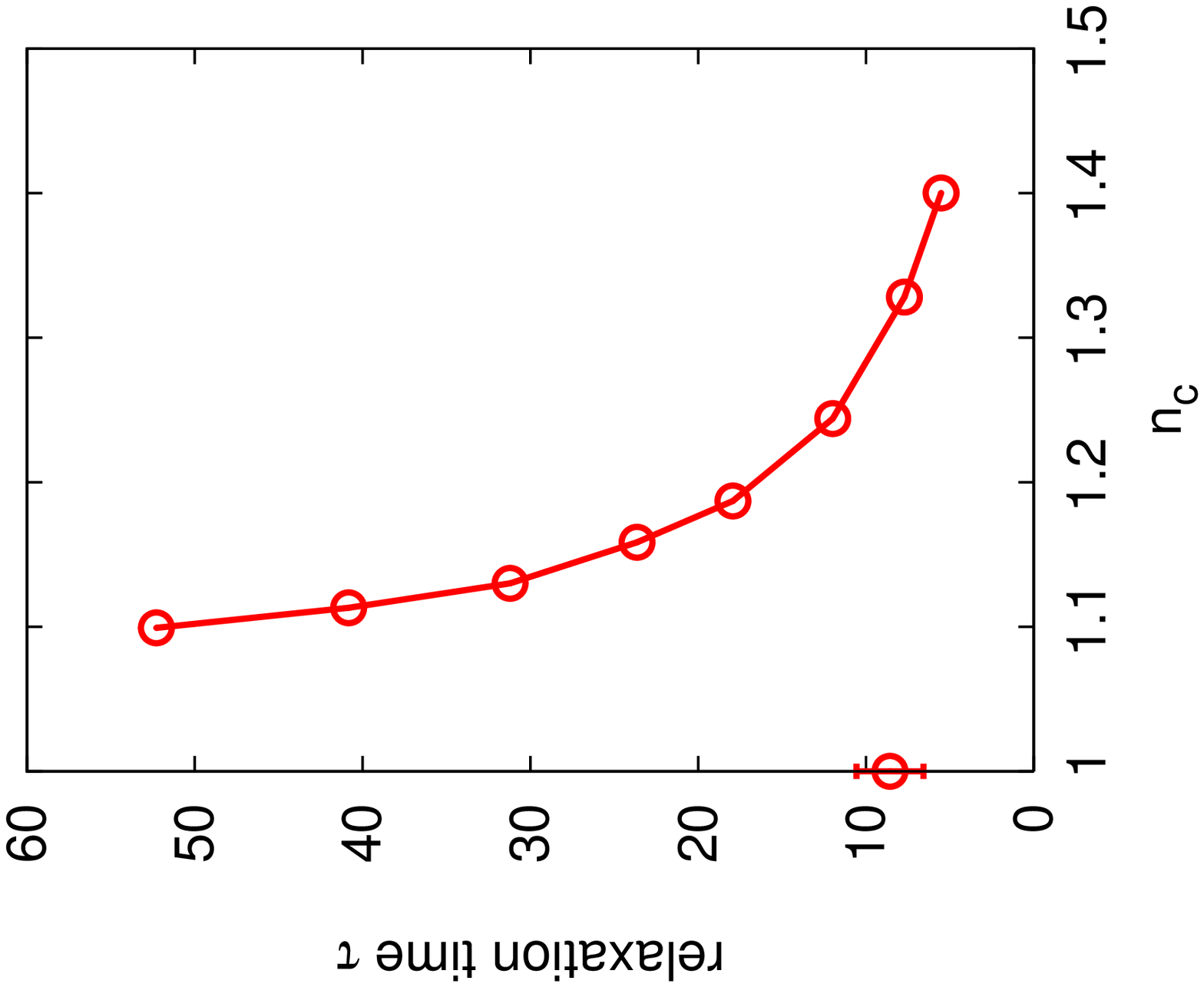}
\includegraphics[angle=-90, width=0.49\linewidth]{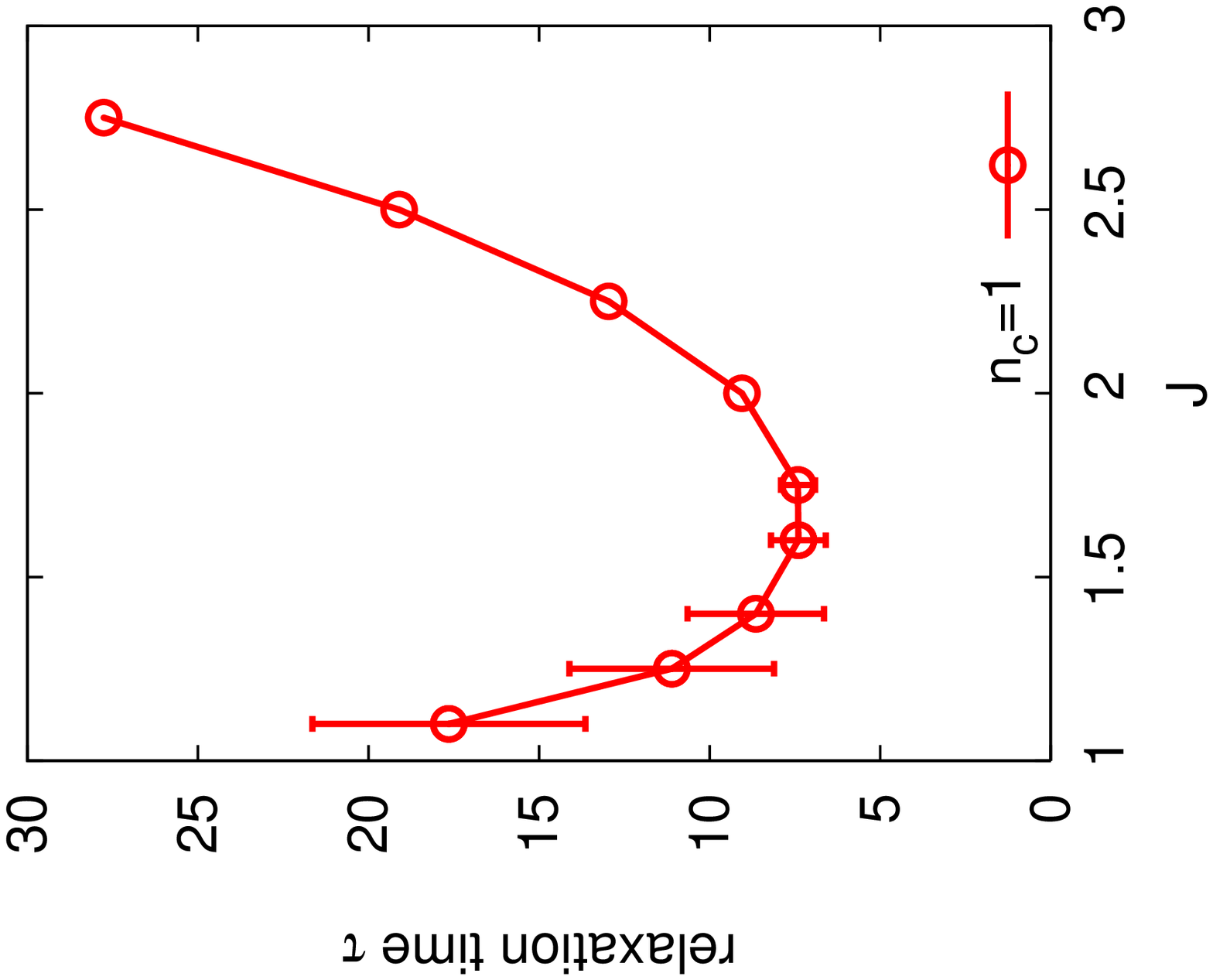}
\caption{ 
Relaxation time $\tau$ for $J=1.5$ as a function of doping (left panel) and at half-filling as a function of $J$ (right panel). 
The initial state is the equilibrium state at $\beta=50$.
The large error bars at small $J$ are due to the appearance of oscillations on top of the exponential relaxation.
}
\label{fig_tau_n}
\end{figure}

\begin{figure*}[t!]
\centering
\includegraphics[angle=-90, width=0.33\linewidth]{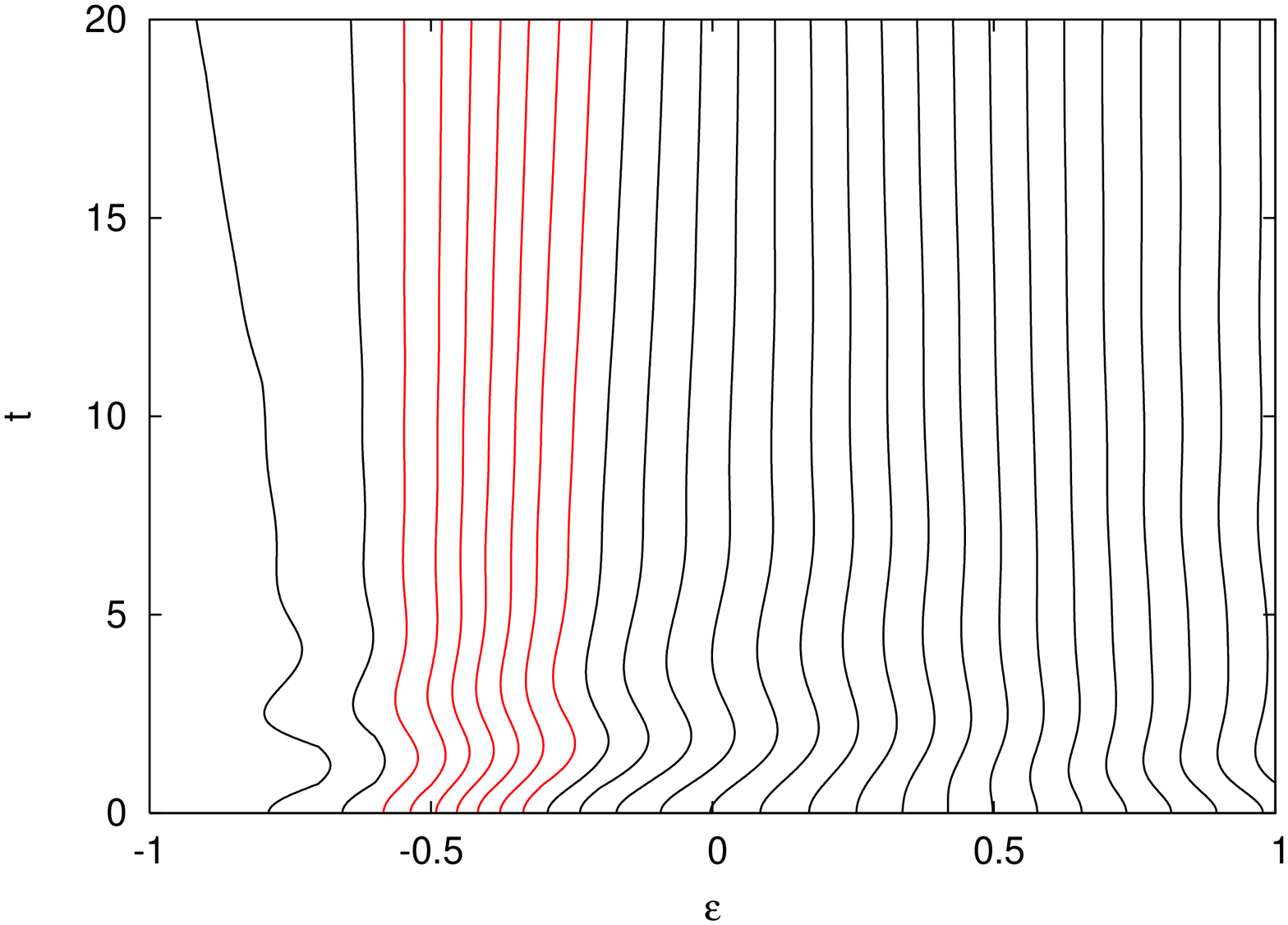}
\includegraphics[angle=-90, width=0.33\linewidth]{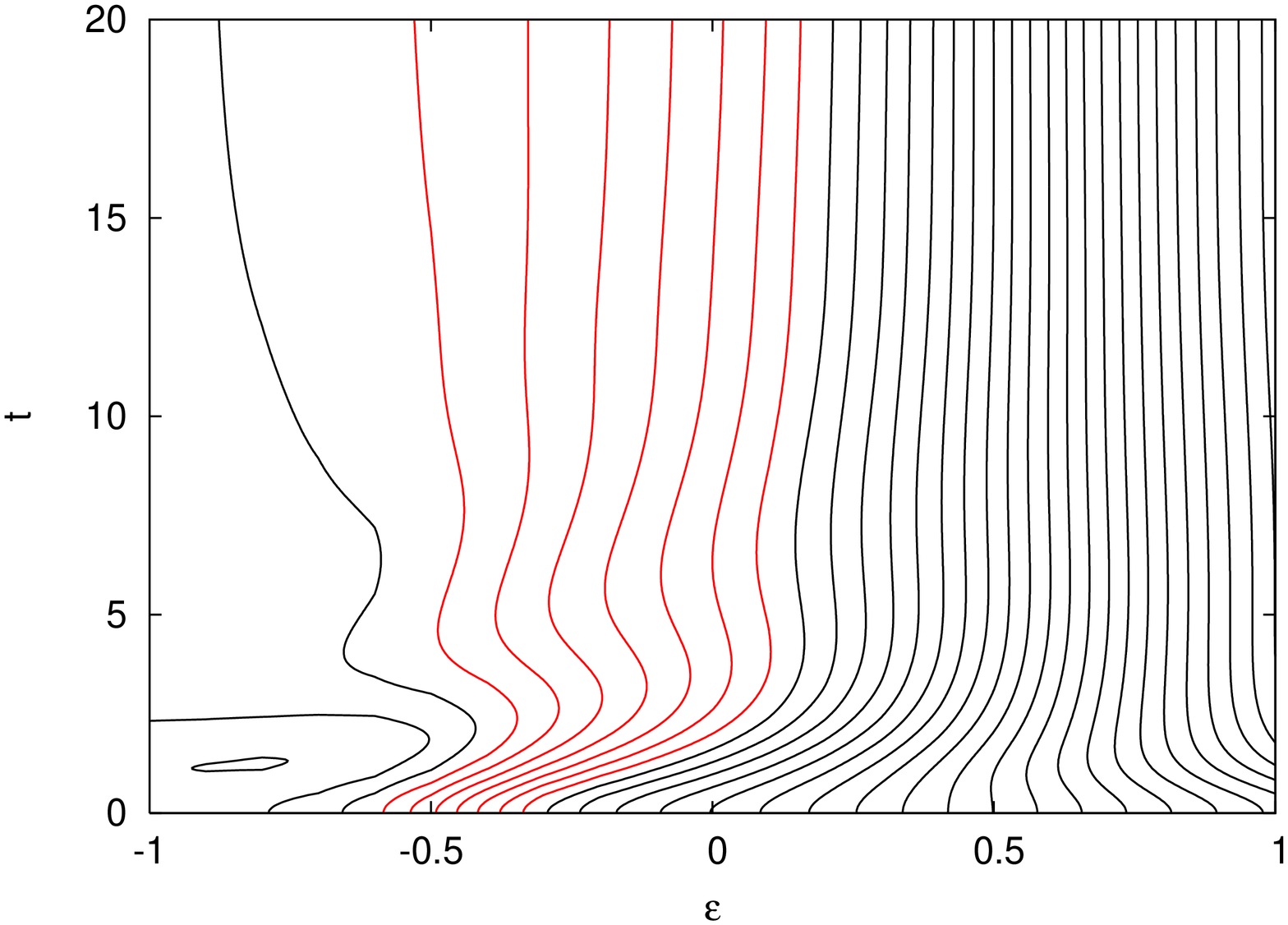}
\includegraphics[angle=-90, width=0.33\linewidth]{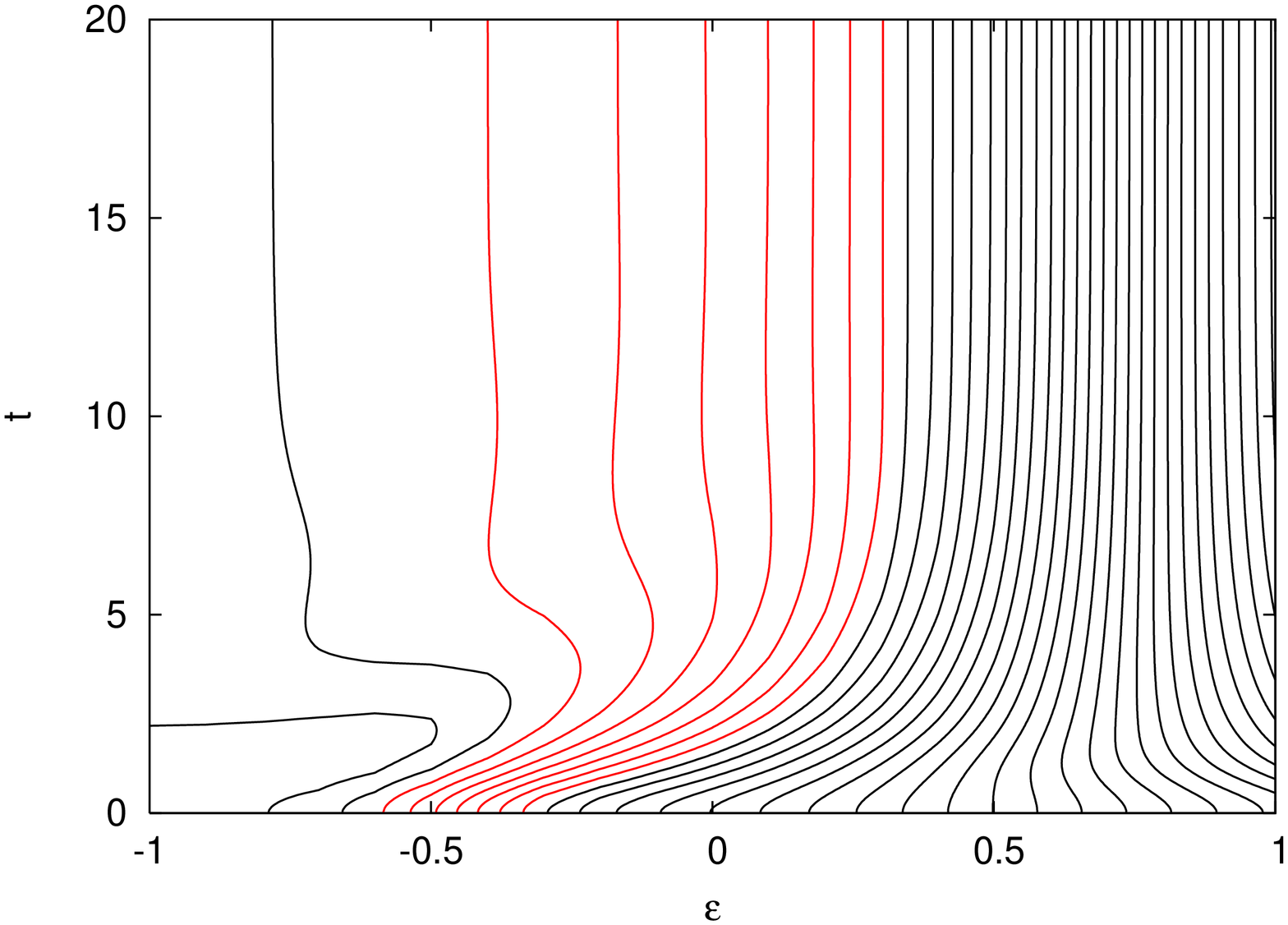}
\includegraphics[angle=-90, width=0.33\linewidth]{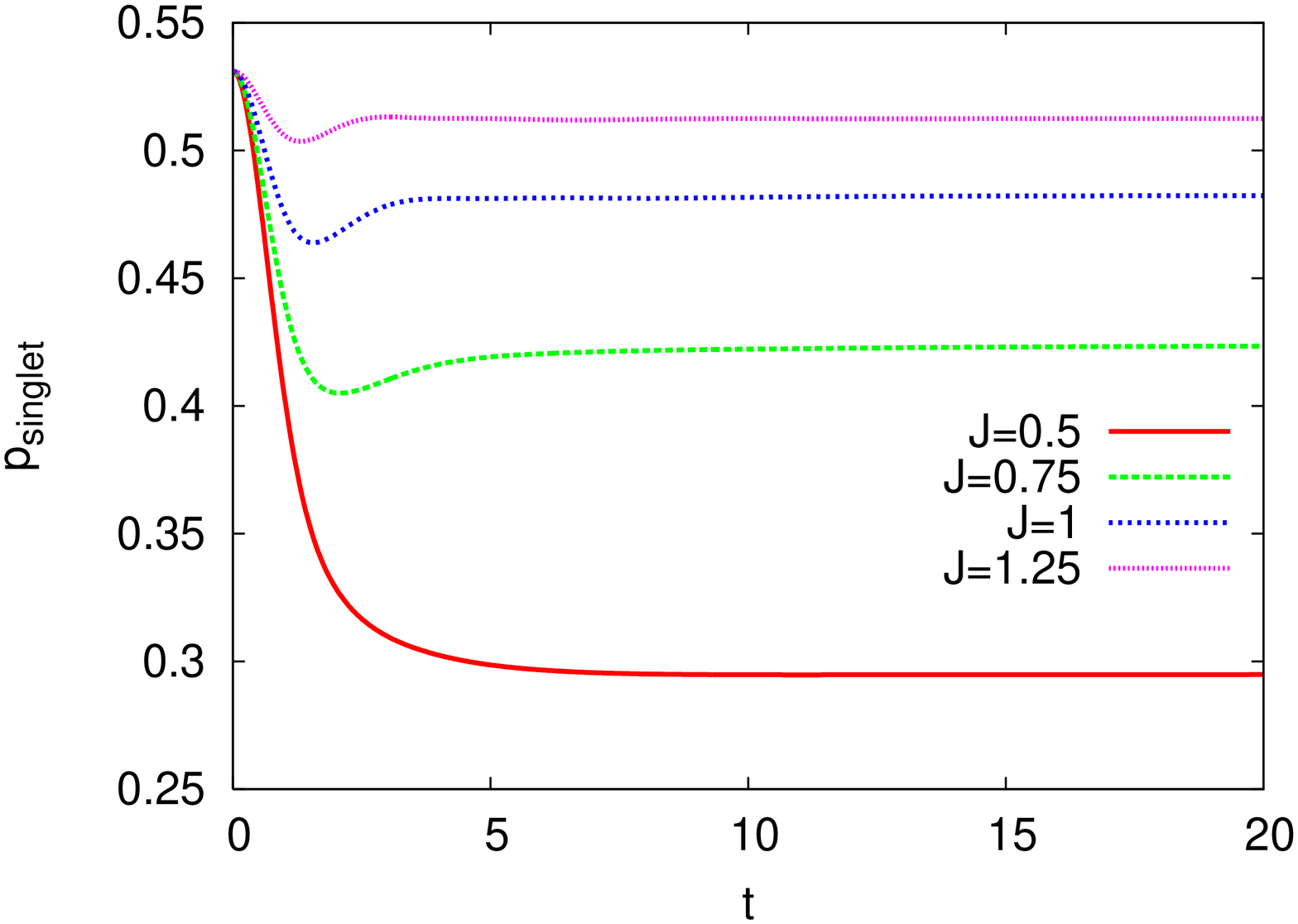}
\includegraphics[angle=-90, width=0.33\linewidth]{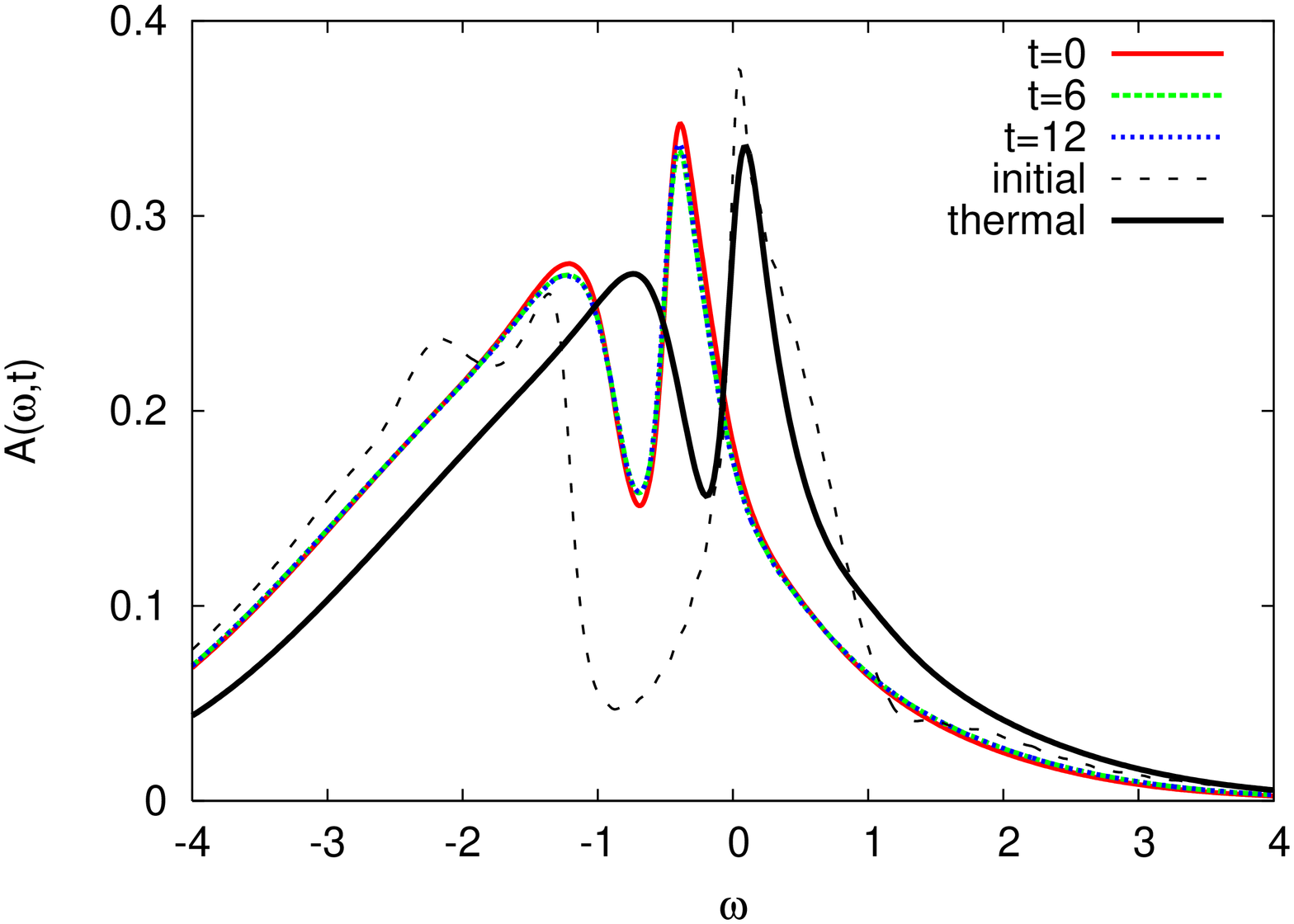}
\includegraphics[angle=-90, width=0.33\linewidth]{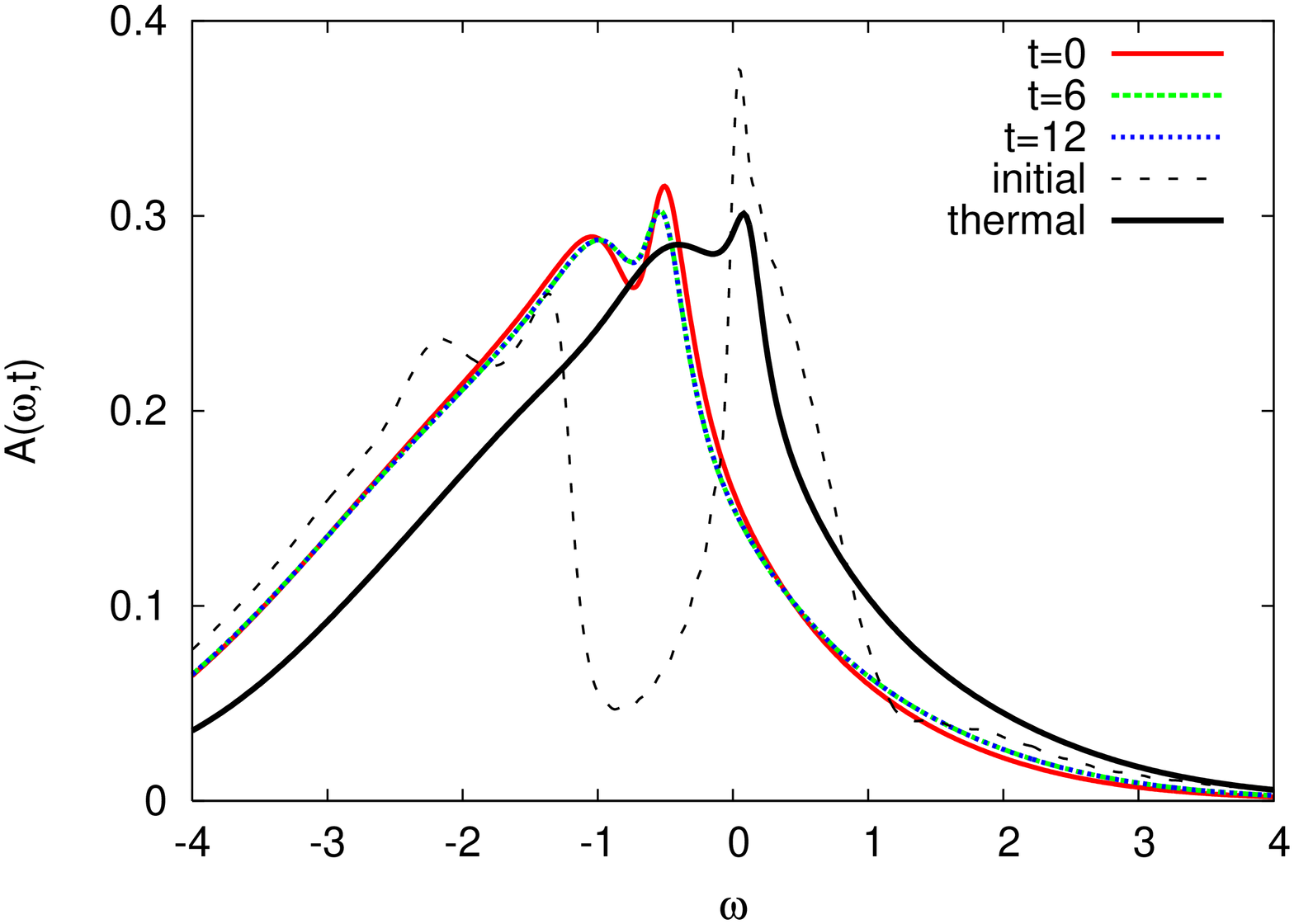}
\caption{Top panels: Time evolution of the distribution function $n(\epsilon,t)$ after a quench from $J=1.5$, $n_c=1.4$, $\beta=50$ to $J=1.25$, $0.75$ and $0.5$ (from left to right). 
The plots show contours of constant $n$, with red contours corresponding to values of $\epsilon$, which in the initial state are associated with the 
large Fermi surface
discontinuity.
Bottom left panel: Time evolution of $p_\text{singlet}$. Middle panel: Time evolution of the spectral function after a quench to $J=0.75$. After thermalization, the system is in the KI phase ($\beta\approx 12.5$). Right panel: Time evolution of the spectral function after a quench to $J=0.5$ (effective temperature $\beta\approx 8.3$, in the LM phase). 
}
\label{fig_contour_quench}
\end{figure*}

The behavior in Fig.~\ref{fig_relax_T} is reminiscent of the relaxation dynamics found after an electric field pulse in the Hubbard model, as one approaches 
the insulator-metal crossover regime from the Mott insulating side.\cite{Eckstein11pump} 
In the doped Kondo lattice model, however, the slow relaxation occurs in the heavy Fermi liquid regime, 
while the KI regime with a deep pseudo-gap is associated with faster relaxation. 

In fact, it is the disappearance of coherent quasi-particles, and not the filling-in of the pseuo-gap, which leads to the faster relaxation. To illustrate this, we plot in Fig.~\ref{fig_relax_vs_resigma} the relaxation time as a function of temperature, and compare this curve to $\text{Re}\Sigma(0)$ and $p_\text{singlet}$. The relaxation time tracks $\text{Re}\Sigma(0)$ 
(associated with the formation of the heavy Fermi liquid 
with a large Fermi surface) 
and not $p_\text{singlet}$ (associated with the opening of the gap).
Furthermore, the relaxation time decreases if the Fermi liquid is weakened, for example by increasing the doping  (left panel of Fig.~\ref{fig_tau_n}).
That increasing doping leads to  a weakening of the heavy Fermi-liquid state is indicated by the evolution 
of $\text{Re}\Sigma(0)$ shown in Fig.~\ref{fig_re_sigma},  which  shows 
that the FL state is formed at lower temperature for $n_c=1.4$ than for $n_c=1.1$, and by the spectral functions in Fig.~\ref{fig_kondo_doped}, 
which show that the Kondo gap gets filled in with increasing doping.  In the exhaustion limit $n_c\rightarrow 2$, the Fermi liquid coherence temperature $T^*$ 
should drop to zero.\cite{Burdin00}

At least in the case $J=1.5$ considered here, the Kondo insulator relaxes much faster than the weakly doped 
FL (about the same relaxation time as in the weakly doped model in the KI regime above $T^*$). For smaller couplings, the relaxation time of the
insulator increases rapidly (right panel of Fig.~\ref{fig_tau_n}), since in the small-$J$ limit the slowest processes are $\sim 1/J$. In the limit of large $J$, the relaxation time grows because it becomes difficult to transform an excitation energy of order $J$ into kinetic energy.

\subsection{Fast melting of the large Fermi surface}
\label{secc}

In this section, we study the dynamical evolution of the heavy Fermi liquid with a  large Fermi surface into a state with small Fermi surface.
This process can easily be triggered in many ways, provided a sufficient amount of energy is injected into the system.
We will suddenly reduce the Kondo coupling $J$, which both leads to a considerable heating of the system
and weakens the local singlets. The two effects combined  are expected to result in a crossover from the 
FL phase into the KI and LM regimes.  
Figure~\ref{fig_contour_quench} shows contour plots of the momentum occupation $n(\epsilon,t)$ after 
a quench from $J=1.5$, $\beta=50$, $n_c=1.4$ to $J_\text{final}=1.25$, $0.75$ and $0.5$. The small  quench ($J_\text{final}\!=\!1.25$) leads to a minor shift and a 
thermal broadening of the 
large Fermi surface
(indicated by the red contour lines). The slow drift of the contour lines indicates that the relaxation time is slow, 
as expected for relaxation processes  within the FL regime. The quench to $J_\text{final}=0.75$ drives the 
system into the KI regime, where Fermi liquid coherence is lost (with a corresponding shift in $\text{Re}\Sigma(0)$), 
but the Kondo gap is still present. The relaxation 
is faster, in accordance with the faster dynamics measured after a weak magnetic field pulse in this regime. 
Finally, the quench to $J_\text{final}=0.5$ brings the system into the LM regime, characterized by a 
small Fermi surface. 
Melting of the large Fermi surface and steepening of the momentum distribution around the location of the  
small Fermi surface happen on the timescale of a few inverse hoppings,  comparable to the relaxation time 
after weak perturbations within the LM phase. We observe no additional bottleneck connected to the destruction 
of the large Fermi surface when hopping and $J$ are comparable in magnitude.

The evolution of the system into the KI and LM phases is also evident from the behavior of $p_\text{singlet}$ and the shift of $\text{Re}\Sigma$.
Despite the heating effect, $p_\text{singlet}$ remains large after quenches to $J=1.25$ and $J=1$, and the crossover into the LM regime is 
evident in the form of a substantial drop of $p_\text{singlet}$ for $J=0.5$ (lower left panel of Fig.~\ref{fig_contour_quench}). 
The shift of $\text{Re}\Sigma$, on the other hand
is evident in the bottom right two panels (corresponding to the quenches to $J=0.75$ and 
$J=0.5$): A time-dependent shift 
of $\text{Re}\Sigma(0)$ is equivalent to a shift of the chemical potential $\mu$. In nonequilibrium DMFT calculations, a sudden shift of $\mu$ by $\Delta\mu$ results in a rigid shift of the spectral function (\ref{A_omega_t}) by $-\Delta\mu$ on the frequency axis. In the present case, the rapid decrease of $\text{Re}\Sigma(0)$ leads to a shift of the quasi-particle peak by $+\Delta(\text{Re}\Sigma(0))<0$. Thermalization of the spectral function is seen to occur approximately on the same time-scale as the momentum distribution function.

\subsection{Formation of the heavy Fermion state upon cooling}

Finally, let us consider the dynamical formation of the FL state out of the LM phase. 
In general, it is not possible to reach the FL phase from the LM phase by a sudden increase 
of $J$, since the strong heating caused by such a quench would destroy the Fermi 
liquid coherence. We will thus approach the problem in a different way, which is at the 
same time closer to possible experiments on heavy-Fermion materials. In such an experiment, 
one might 
rapidly destroy the FL phase by a strong excitation (see Sec.~\ref{secc}), and monitor 
its reappearance out of the excited LM phase while energy is  dissipated from  the electronic system to the 
environment. As long as the intrinsic thermalization times of the system are fast compared 
to this dissipation time, the system will evolve through
a sequence of equilibrium states of 
decreasing temperature, at a rate that is set by the coupling to the environment.  
However, when the thermalization time becomes long, the system will fall out of 
equilibrium, or experience a slowdown of the cooling dynamics. Since our investigation 
in Sec.~\ref{secb} has demonstrated a large increase of the thermalization times in the low-tempeature  
FL phase, one might  
expect the cooling dynamics in the doped Kondo lattice to reveal such
a nontrivial time evolution.
In the following, we will excite the system with a strong magnetic field 
pulse (\ref{hpulse}), and look at the subsequent dynamics in the presence of an additional particle 
reservoir at fixed temperature. 

To model dissipation 
of energy to other degrees of freedom 
we couple  a thermal particle reservoir with inverse temperature $\beta$
locally to each site of the lattice. Technically, this corresponds to a change of the DMFT self-consistency condition from Eq.~(\ref{self-consistency}) to 
\begin{equation}
\Delta(t,t')=v^2 G_c(t,t') + \Delta_\beta(t,t'),
\label{self-consistency-with-bath}
\end{equation}
where the hybridization function of the reservoir is given by 
\begin{equation}
\Delta_\beta(t,t')=-i\int d\epsilon \gamma(\epsilon) [\Theta_\mathcal{C}(t,t')-f_\beta(\epsilon)]e^{-i\epsilon (t-t')}, 
\end{equation}
with $f_\beta(\epsilon)$ the Fermi function (see Appendix). 
A similar set-up has been previously considered in studies of the nonequilibrium properties of the Falikov-Kimball\cite{Tsuji2009} 
and Hubbard models.\cite{Amaricci2011} For $\gamma(\epsilon)$ we use a semi-elliptic function with  bandwidth $16$ 
and amplitude $\gamma(0)=\gamma$. 

\begin{figure}[ht]
\centering
\includegraphics[angle=-90, width=0.495\linewidth]{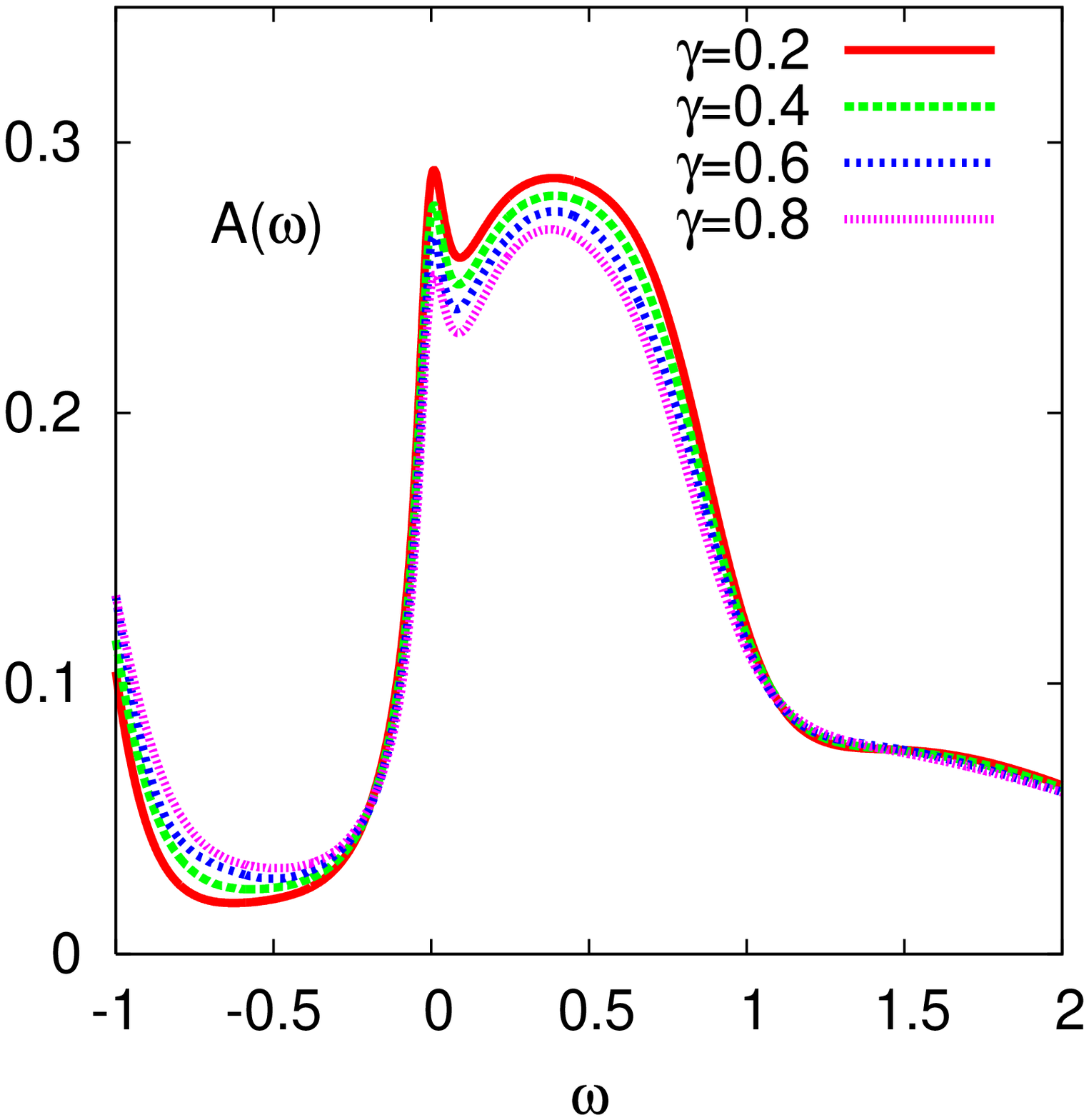}
\includegraphics[angle=-90, width=0.495\linewidth]{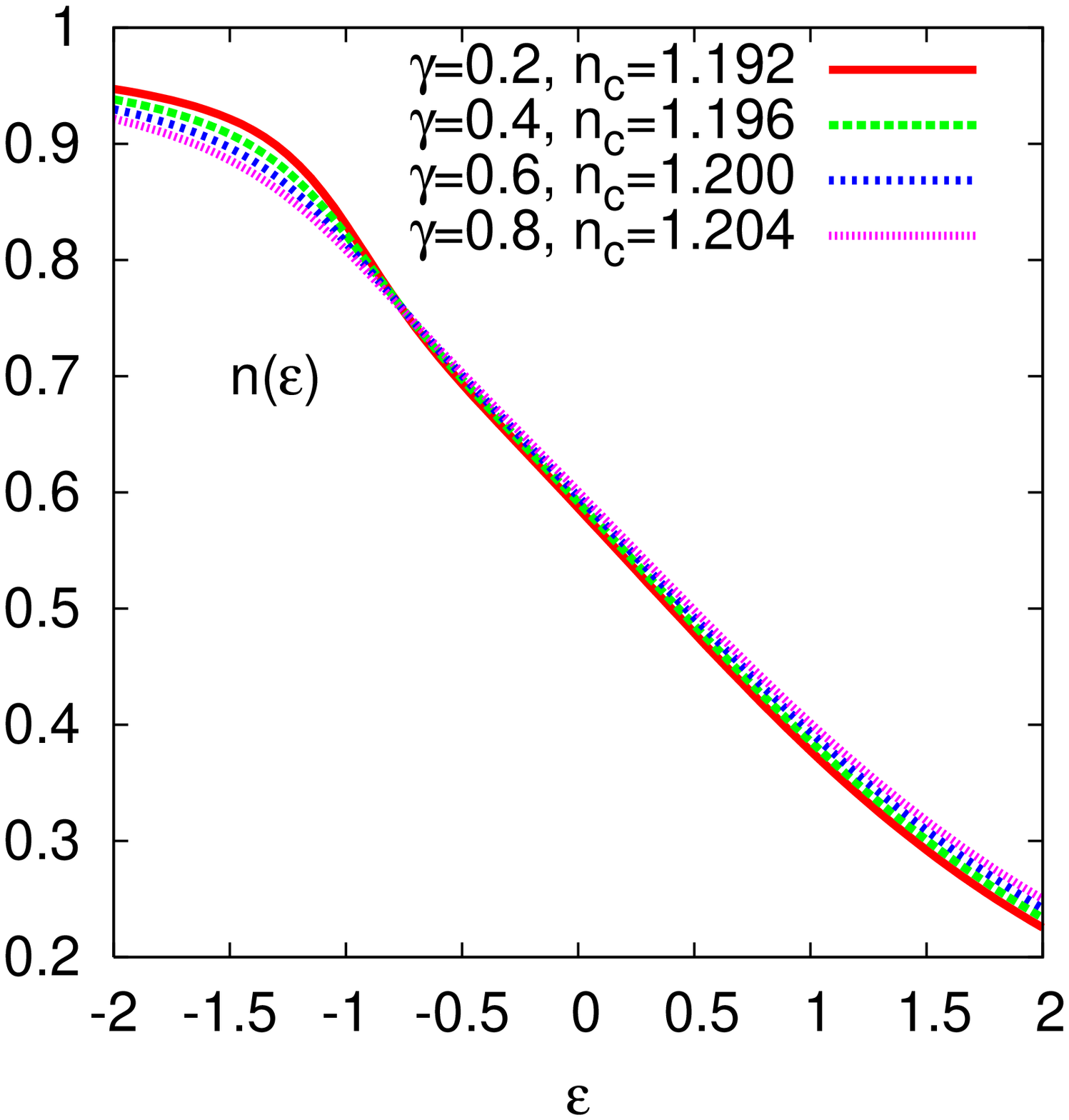}
\includegraphics[angle=-90, width=0.495\linewidth]{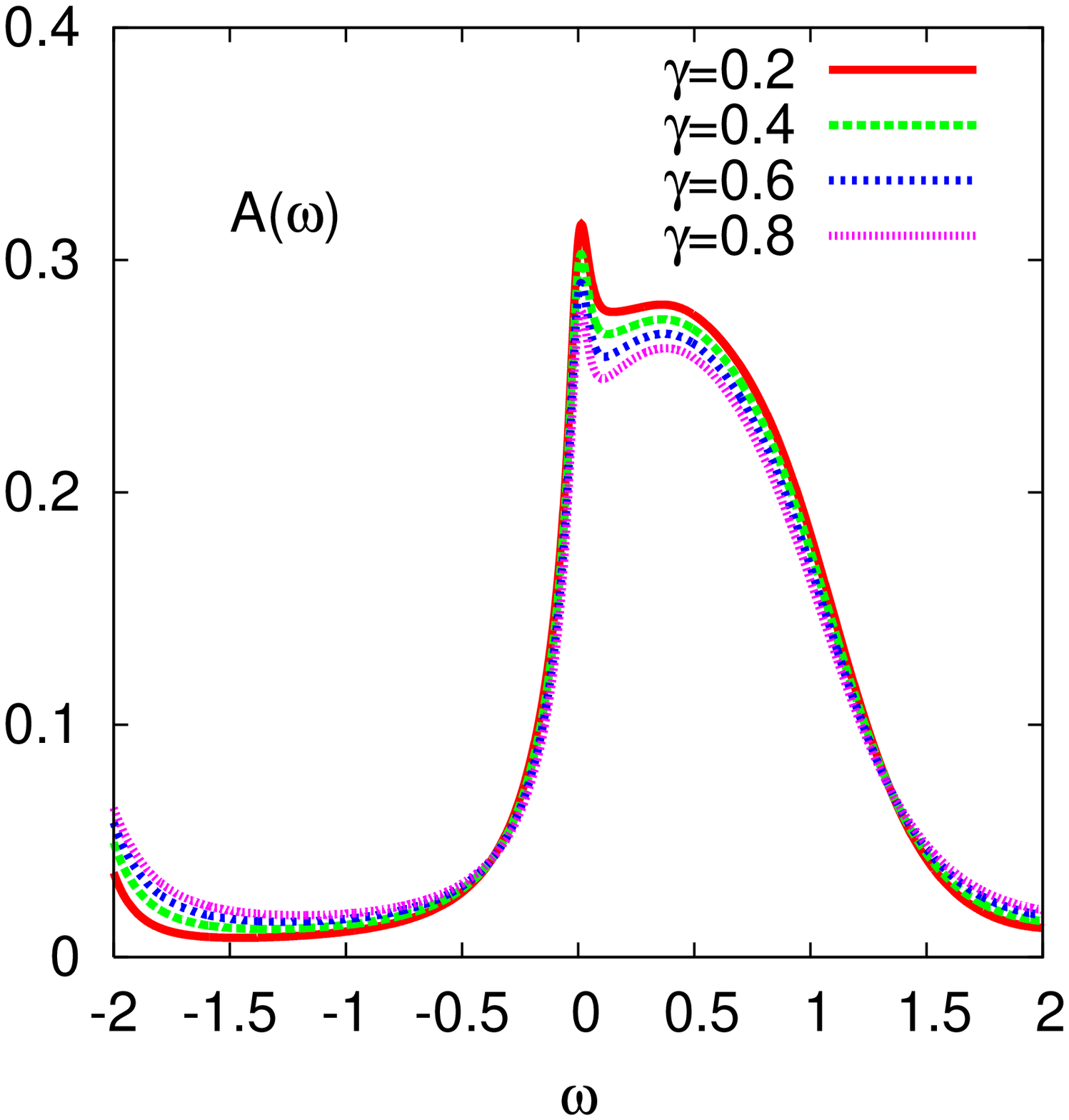}
\includegraphics[angle=-90, width=0.495\linewidth]{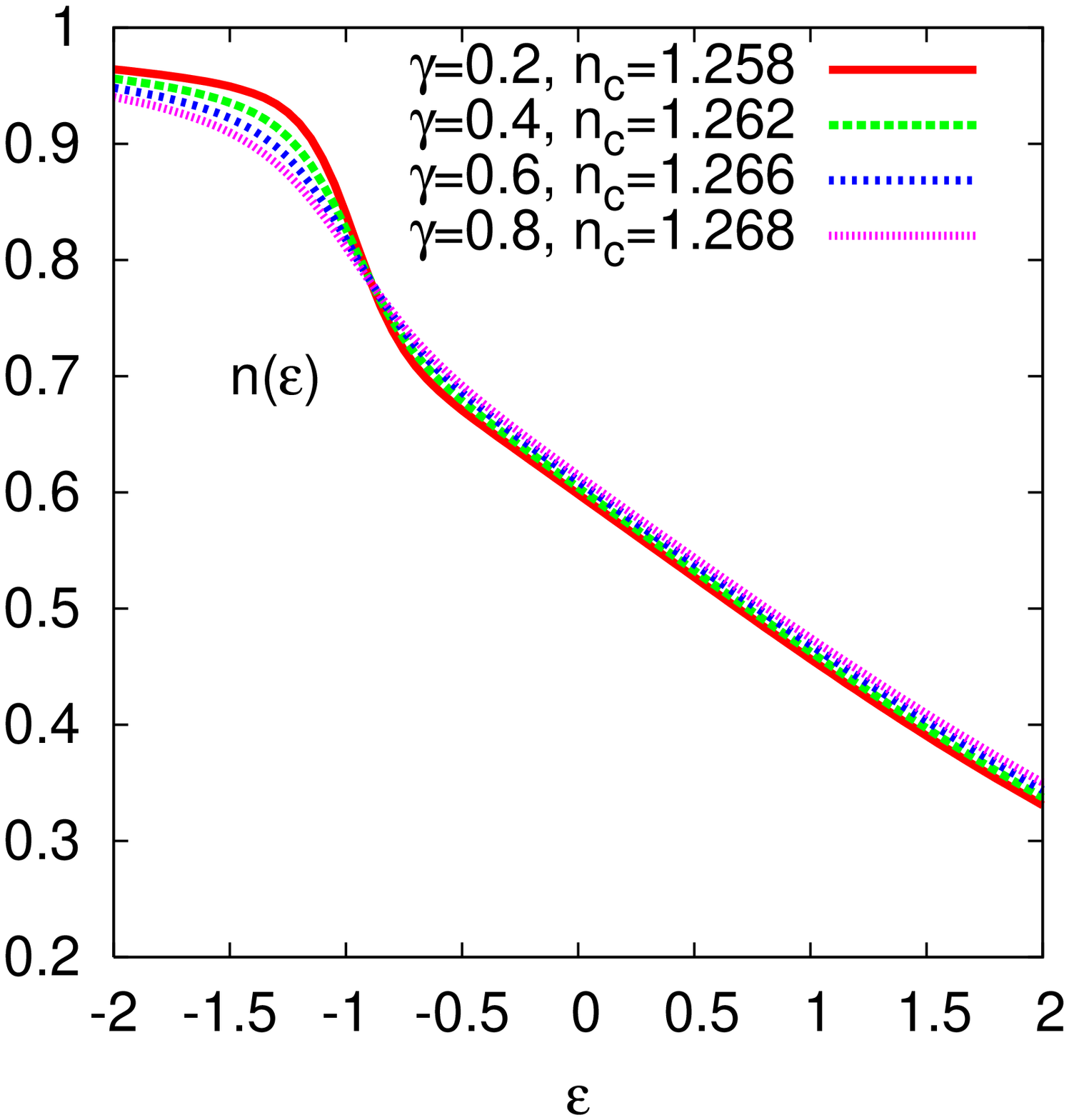}
\caption{Spectral functions $A(\omega)$ (left panels) and distribution functions $n(\epsilon)$ (right panels) for different couplings $\gamma$ to a heat bath with $\beta=50$. Top panels: $J=1.5$. Bottom panels: $J=2.5$. 
}
\label{a_equilibrium_j1.5}
\end{figure}

\begin{figure}[ht]
\centering
\includegraphics[angle=-90, width=0.495\linewidth]{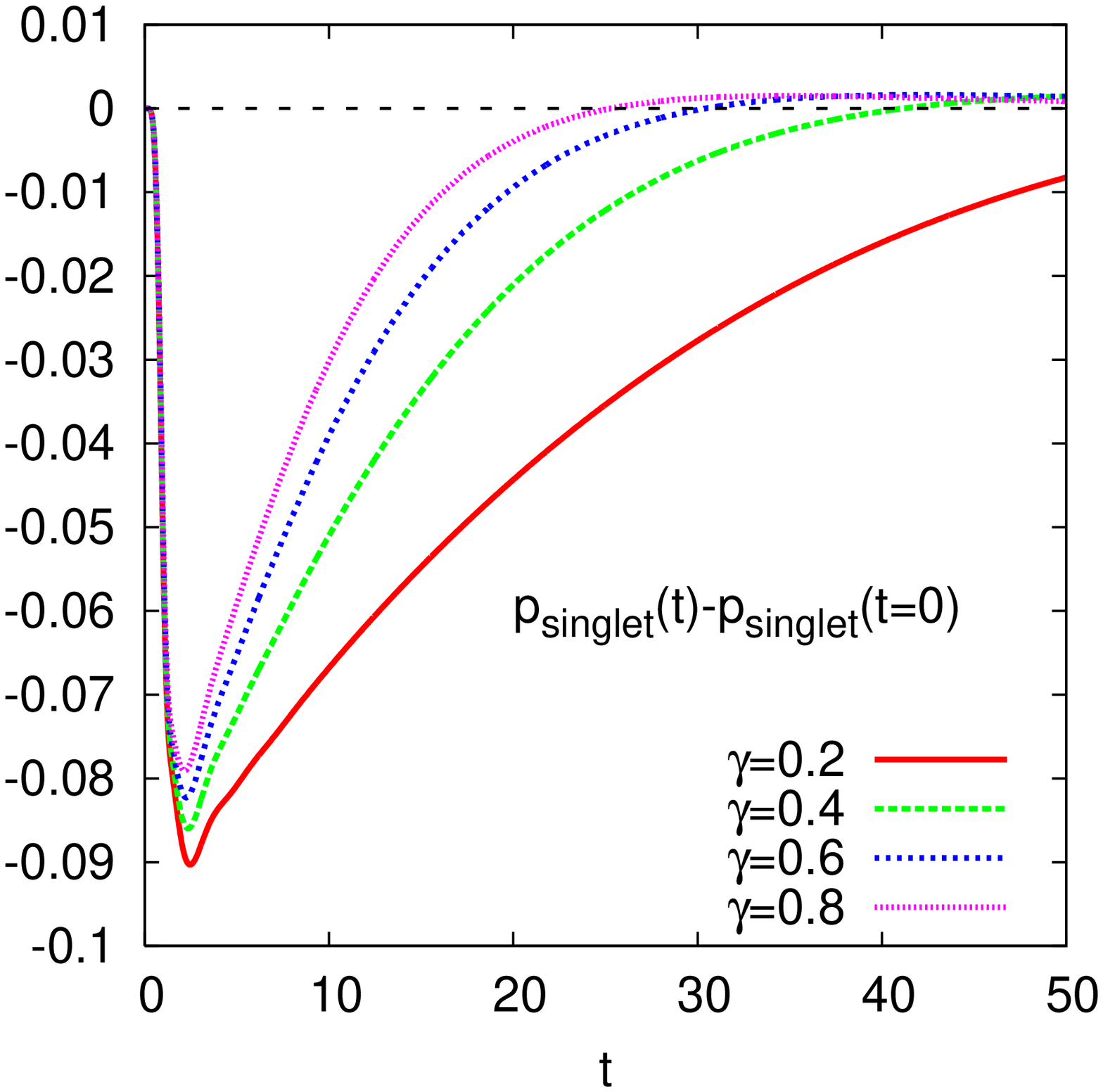}
\includegraphics[angle=-90, width=0.495\linewidth]{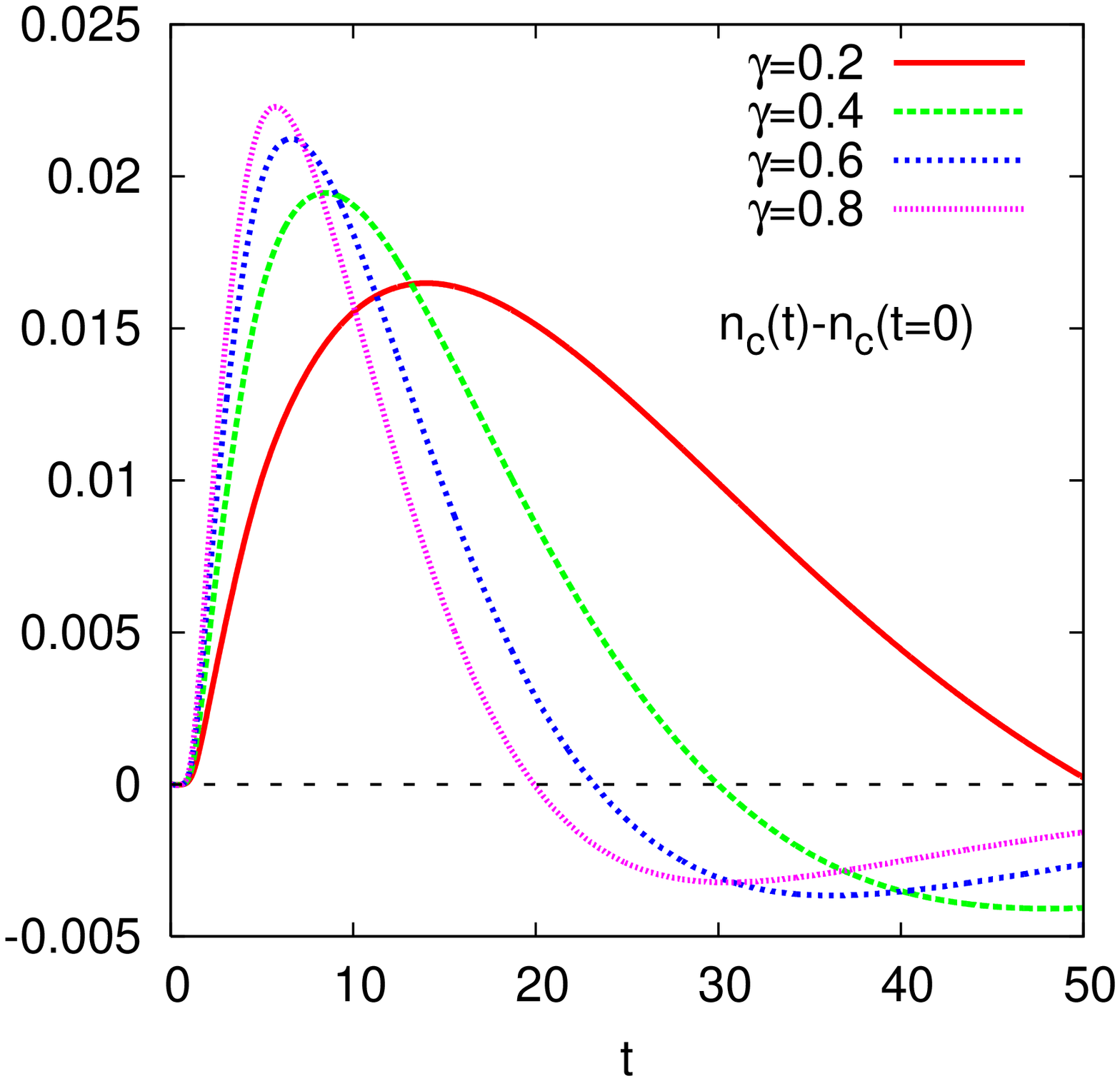}
\includegraphics[angle=-90, width=0.495\linewidth]{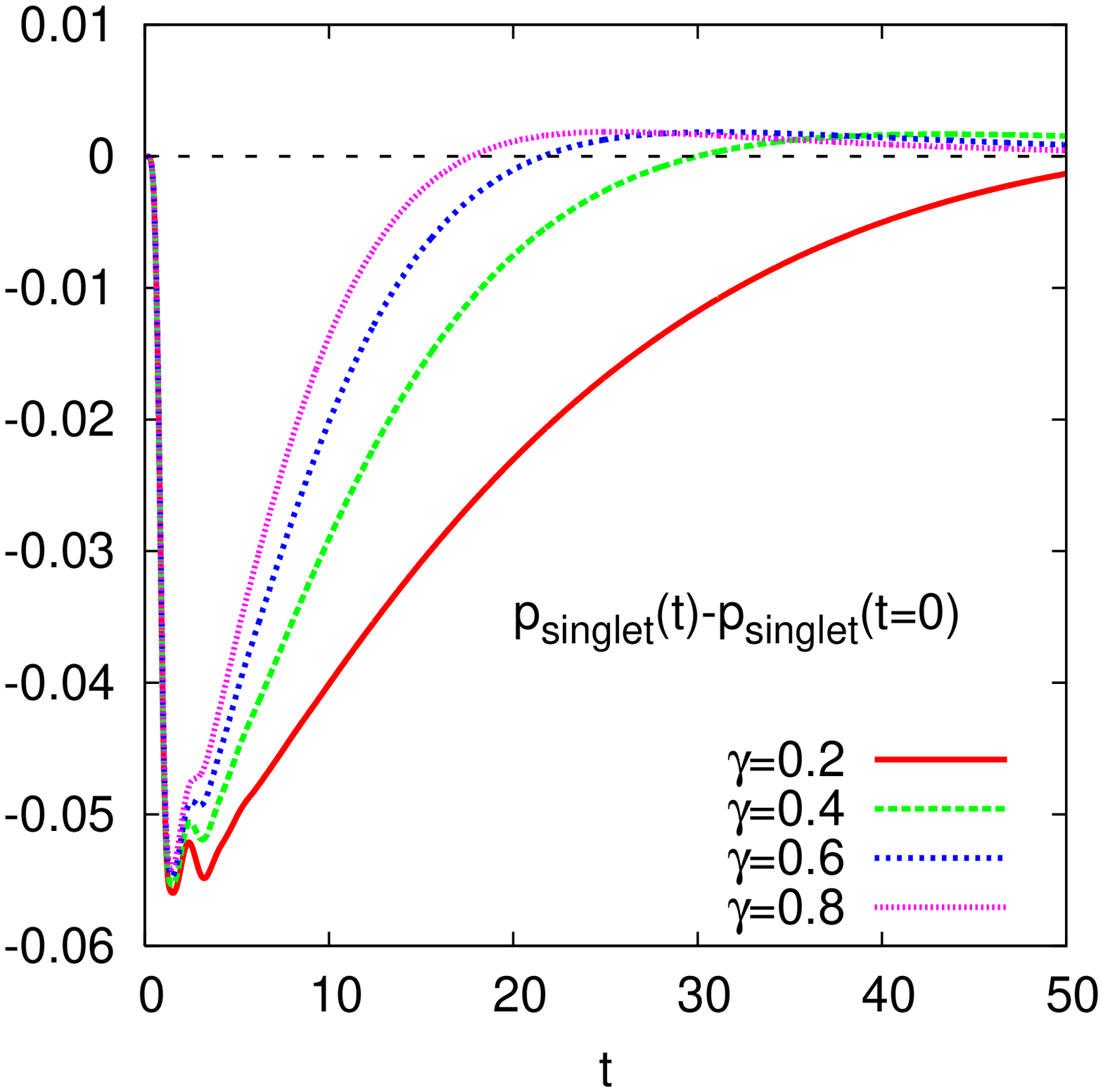}
\includegraphics[angle=-90, width=0.495\linewidth]{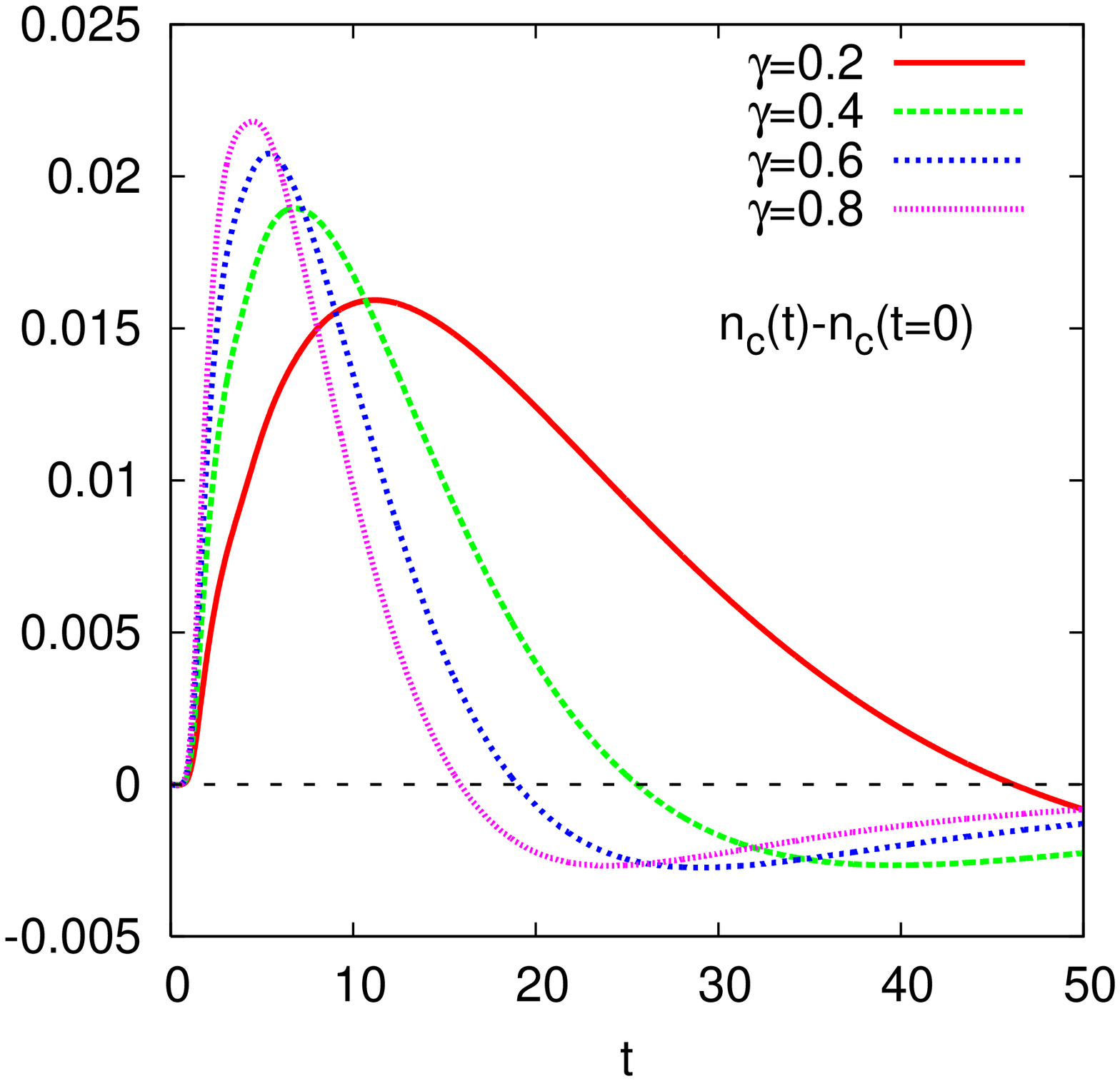}
\caption{
Time-evolution of the probability of the singlet state (left panels) and the density (right panels) for indicated couplings $\gamma$ to the bath with $\beta=50$. The perturbation is a magnetic field pulse of amplitude $h_x=0.5$ and duration $t_\text{max}=1.5$. 
A larger coupling $\gamma$ leads to a faster relaxation back to the thermal value. 
The top panels show data for $J=1.5$ and the bottom panels for $J=2.5$.
}
\label{timeev_j1.5}
\end{figure}

In the following we will focus on values of $\gamma$ which are large enough to allow for a fast energy dissipation, but
still so small that the equilibrium properties of the system remain qualitatively unchanged with respect to the isolated system.
We will therefore first study 
the effect of the bath (with inverse temperature $\beta=50$) on the spectral function and momentum distribution function in equilibrium. 
As shown in the left panels of Fig.~\ref{a_equilibrium_j1.5}, a stronger coupling $\gamma$ reduces the 
peak near $\omega=0$ in the spectral function, 
which is associated with the heavy quasi-particle band.
Spectral weight is added to the gap region.
As a result, the step-like feature in the 
distribution function $n(\epsilon)$ (near $\epsilon=-1$) gets smeared out and for $J=1.5$, $n_c\approx 1.20$ it is hardly evident anymore 
for $\gamma\gtrsim 0.4$ (right panels). To obtain a more prominent large Fermi 
surface, but still keep a strong coupling $\gamma$ to the environment, we 
also consider data for $J=2.5$ and $n_c\approx 1.26$ (bottom panels). 
While the average density is affected by the coupling to the bath, 
this effect is comparatively small: $n_c$ increases only 
by about one percent for a coupling strength $\gamma=0.8$.

The time evolution of the system after 
an intense magnetic field pulse 
with strength $h_x=0.5$ and duration $t_\text{max}=1.5$ 
is plotted in Fig.~\ref{timeev_j1.5}. The left panels show the weight of the average occupation 
of the singlet state, and the right panels show the density. The larger the coupling $\gamma$, the faster these observables relax back to approximately the thermal 
value:
If one defines a relaxation timescale $\tau_{0.002}$ for this initial fast relaxation as 
the time by which $p_\text{singlet}$ reaches its thermalized value up to within $\pm 0.002$,
one finds that these relaxation times $\tau_{0.002}$ scale approximately linearly 
with $1/\gamma$ 
( $J=1.5$: $\tau_{0.002}=62$, $36$, $26.5$, $22$ for $\gamma=0.2,0.4,0.6,0.8$, respectively; 
$J=2.5$: $\tau_{0.002}=48$, $26$, $19$, $15.5$ for $\gamma=0.2,0.4,0.6,0.8$ ).

\begin{figure*}[t]
\centering
\includegraphics[angle=-90, width=0.33\linewidth]{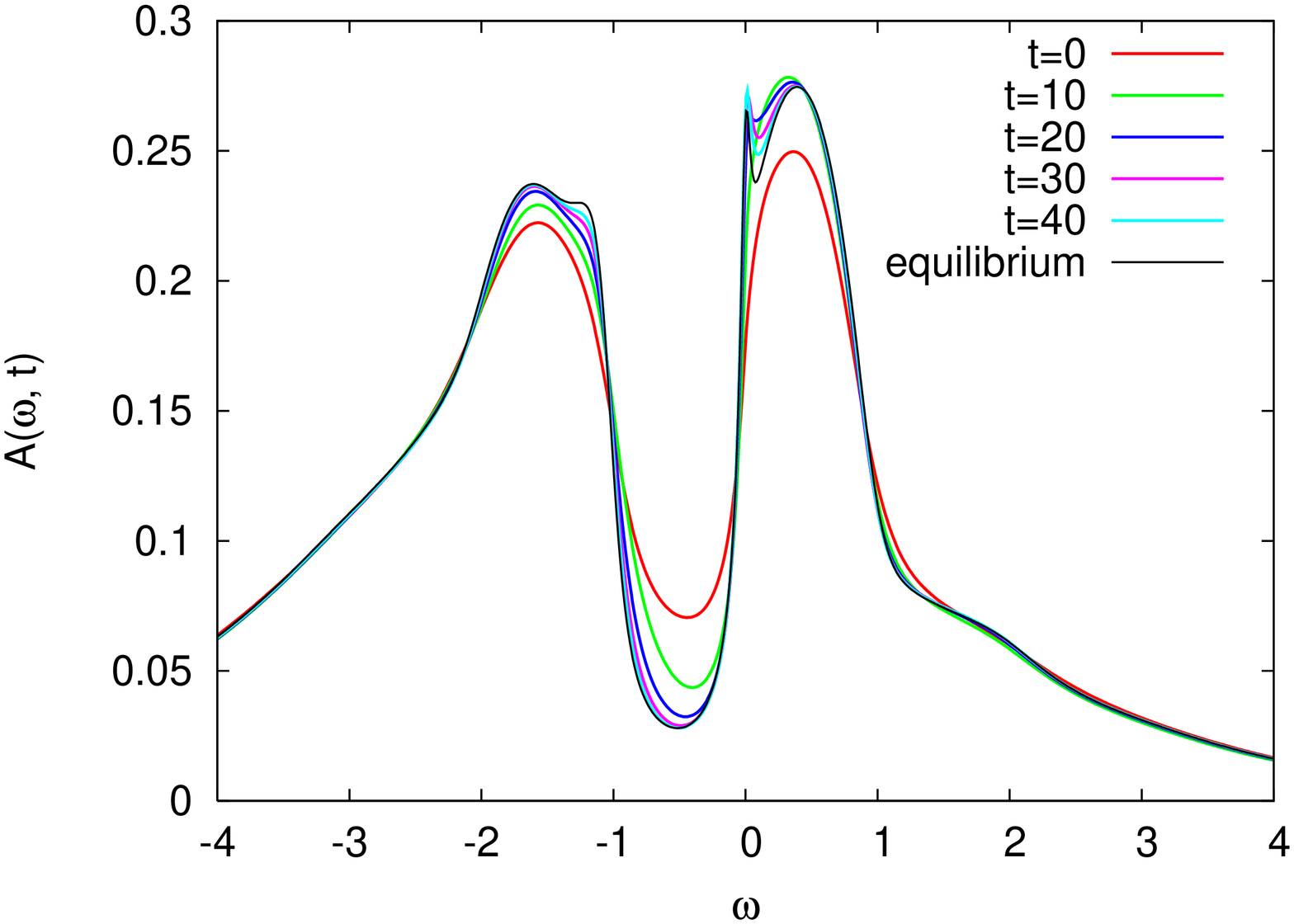}
\includegraphics[angle=-90, width=0.33\linewidth]{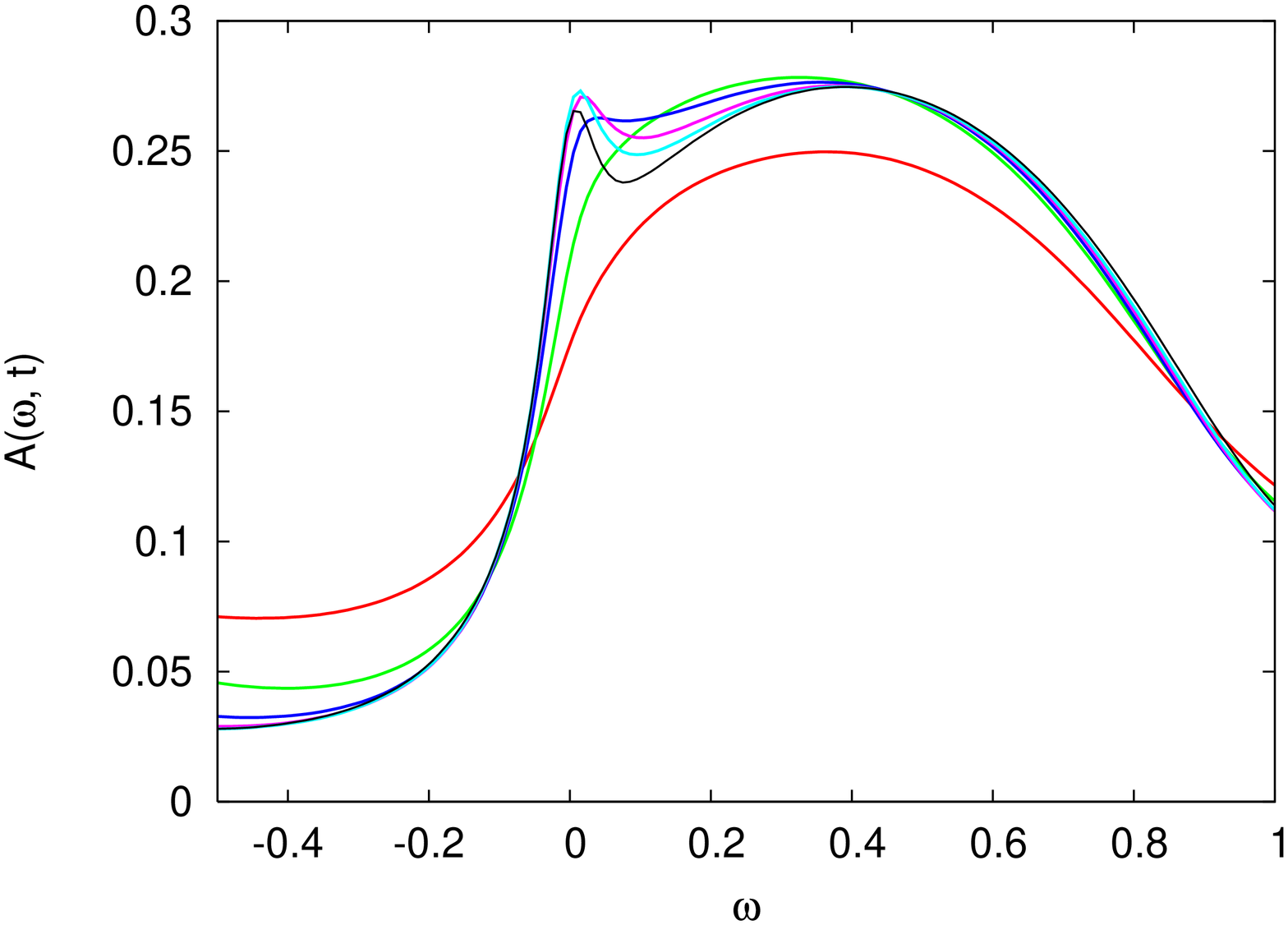} 
\includegraphics[angle=-90, width=0.33\linewidth]{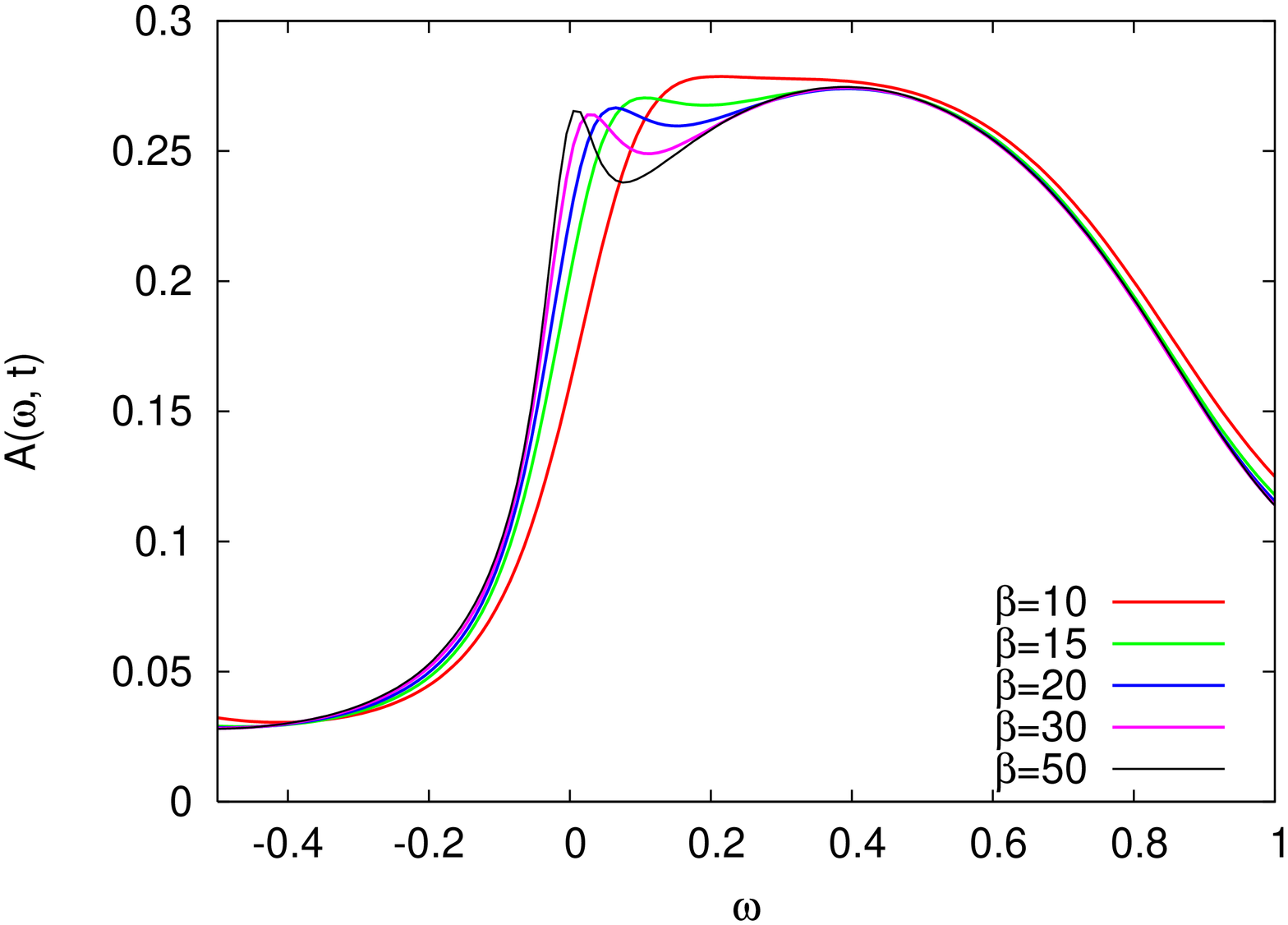} 
\includegraphics[angle=-90, width=0.33\linewidth]{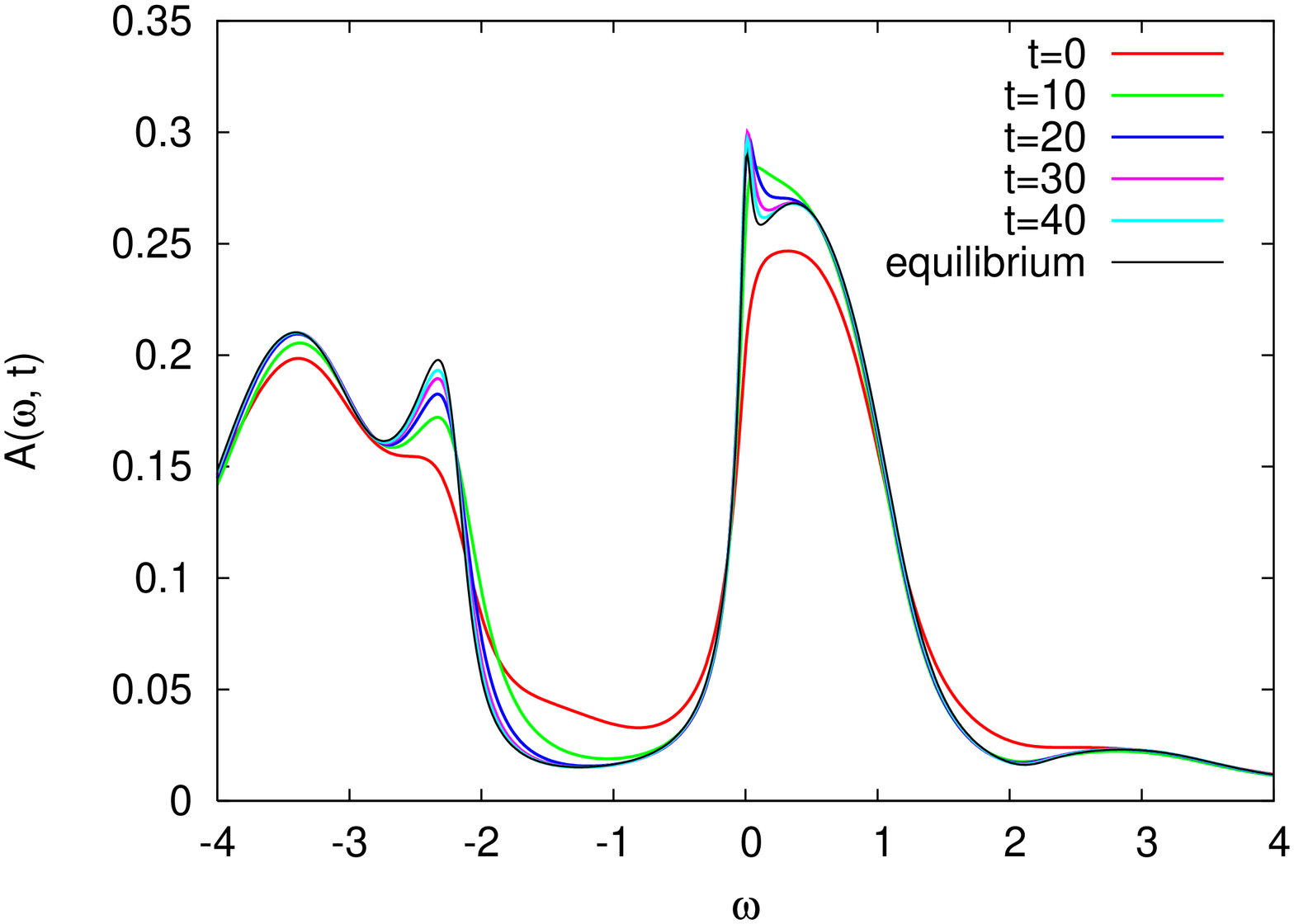}
\includegraphics[angle=-90, width=0.33\linewidth]{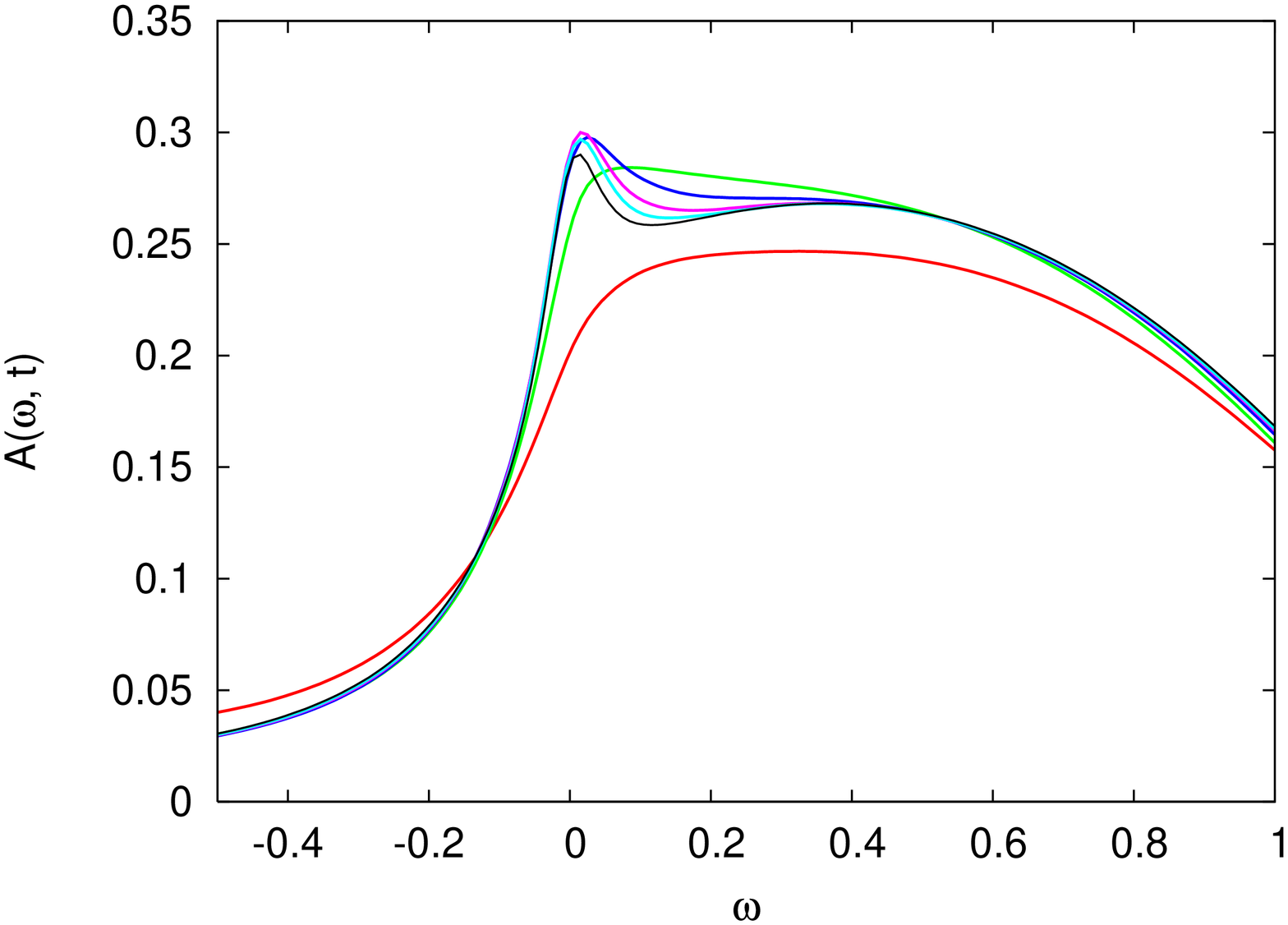}
\includegraphics[angle=-90, width=0.33\linewidth]{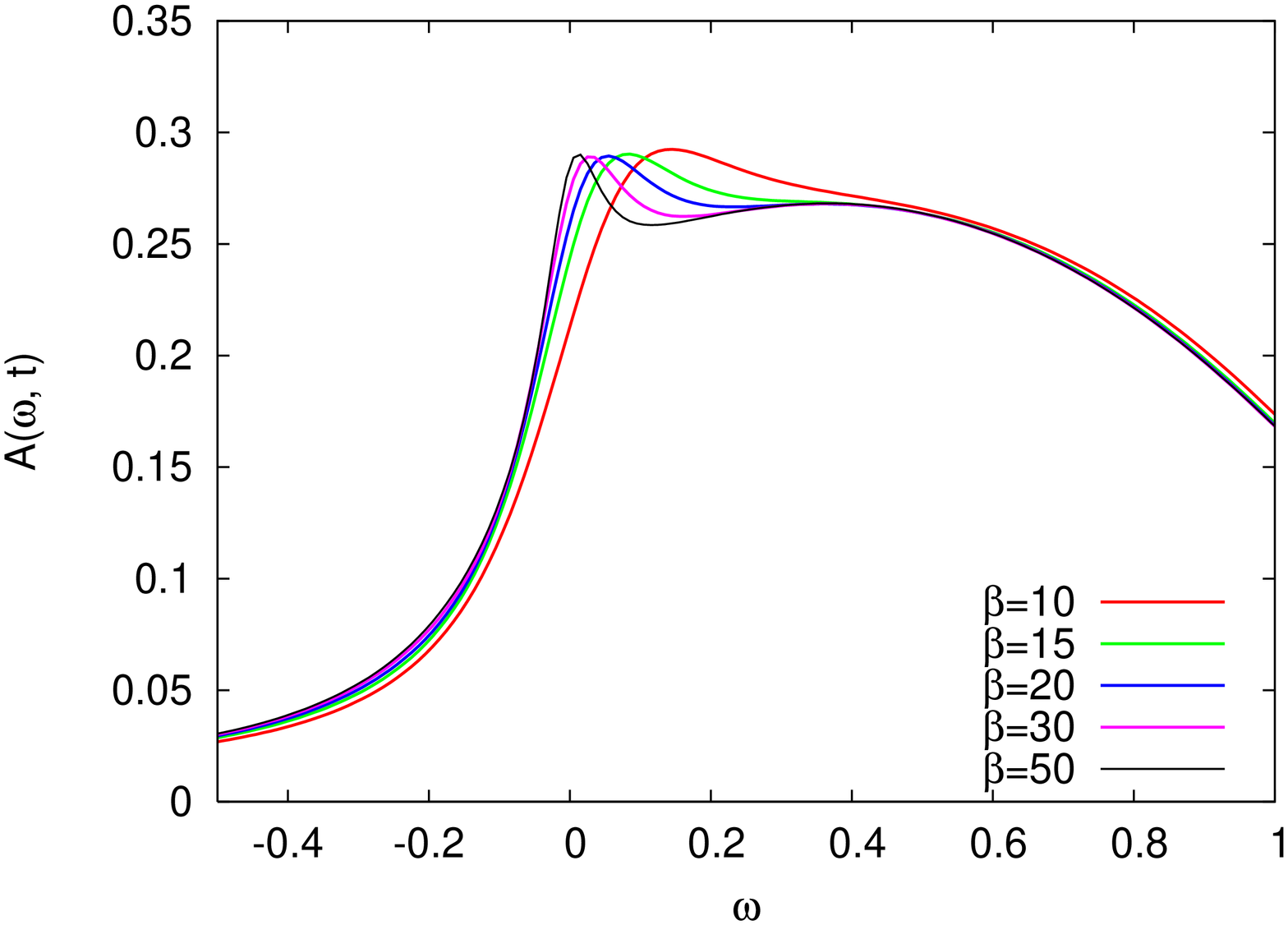}
\caption{Time-evolution of  
the spectral function 
$A(\omega, t)$ for $\gamma=0.6$, $J=1.5$ (top panel) and $J=2.5$ (bottom panel). The left panels show the time-evolution 
on a wide frequency scale and the middle panels a zoom of the quasi-particle peak. For comparison, the right hand panels 
illustrate the evolution of the equilibrium spectral function with temperature.   
}
\label{spectra_j1.5}
\end{figure*}

The fast initial relaxation leads to an 
overshooting of 
the singlet occupation and density, and 
it is followed 
by a slower convergence to the true steady state. 
The fast 
dynamics can be identified with the formation of the  
Kondo gap (more precisely, a pseudo-gap), 
while the slow relaxation is related to 
the appearance of the heavy quasi-particle band.  
To illustrate this fact, 
we plot in Fig.~\ref{spectra_j1.5} the time-dependent 
spectral function [Eq.~(\ref{A_omega_t})]
for several times after the pulse. 
\cite{footnote01}
For $J=1.5$, $\gamma=0.6$ (top panels), first indications of a feature near $\omega=0$ appear around $t=20$ (which is the time needed for 
$p_\text{singlet}$ to relax back to roughly the thermal value), with a well-formed peak and a fully established gap around $t=30$. 
However, the comparison with the equilibrium spectral function (black curve) shows that the amplitude of this peak is enhanced compared to the  
equilibrium 
peak (Fig.~\ref{spectra_j1.5}, middle panels), and the recovery of the latter takes much longer (the thermal curve is not reached for $t\approx 40$, 
which is the largest time for which we can compute well-resolved spectra). 
The observed long times needed to restore the 
heavy quasi-particle band are consistent with the relaxation times $\tau\approx 50$ measured for weak perturbations of the 
FL (Fig.~\ref{fig_relax_vs_resigma}).\cite{footnote02}
Together with the corresponding feature at the lower gap edge, the 
formation of this coherent quasiparticle peak constitutes the bottleneck in the relaxation process. 

\begin{figure}[ht]
\centering
\includegraphics[angle=-90, width=0.9\linewidth]{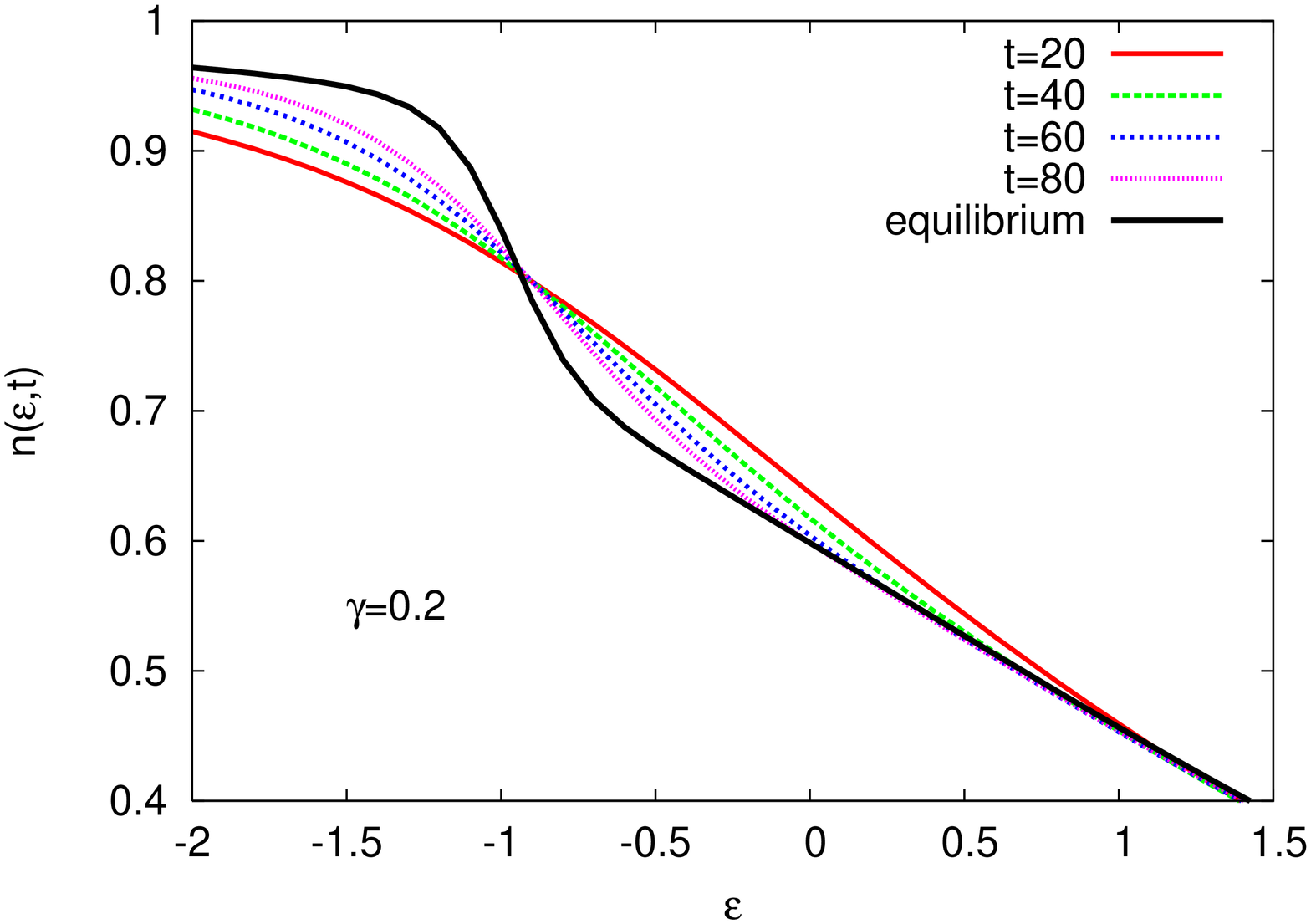}
\includegraphics[angle=-90, width=0.9\linewidth]{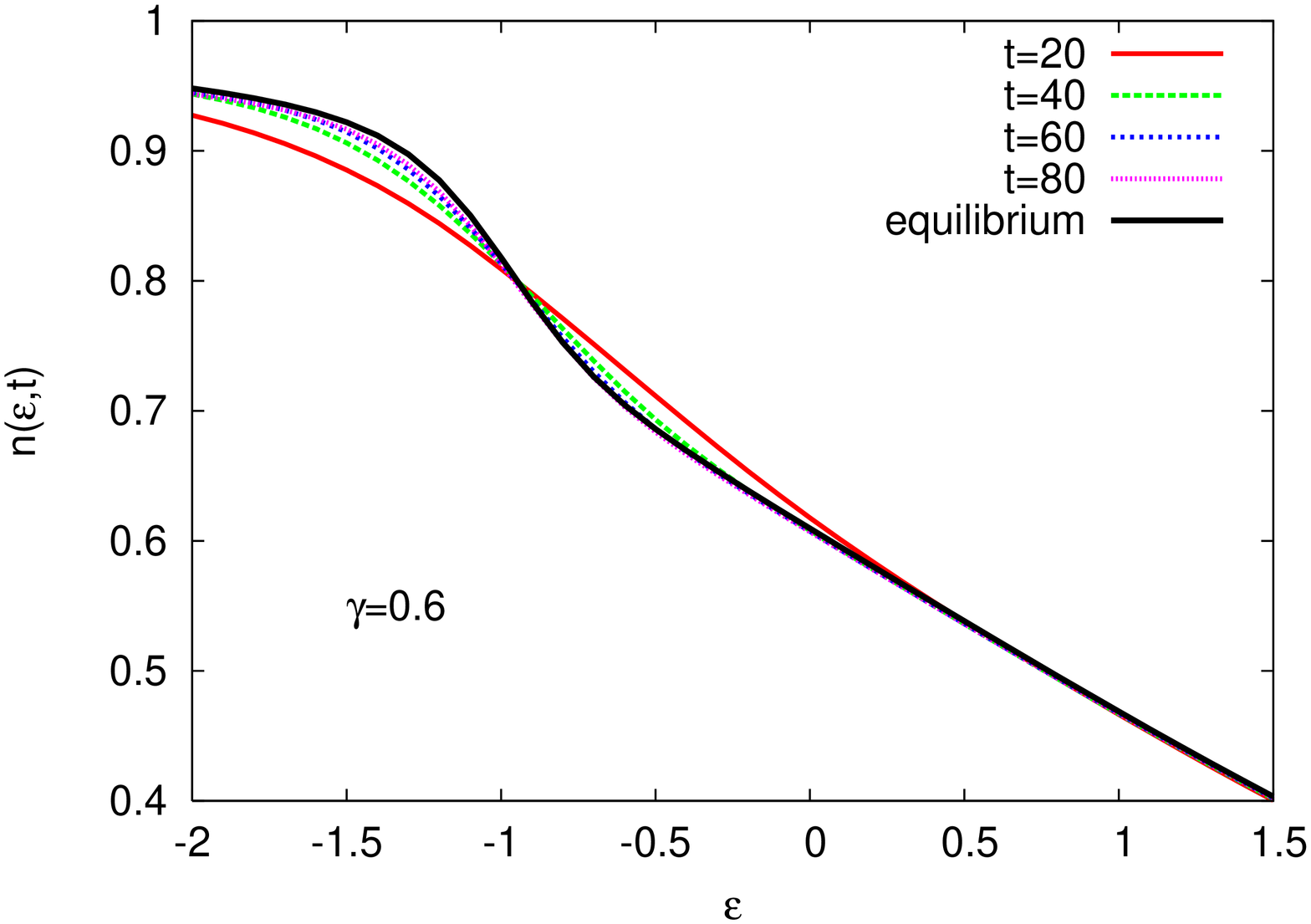}
\caption{$J=2.5$. Time-evolution of the 
momentum distribution function 
for couplings $\gamma=0.2$ (top) and $\gamma=0.6$ (bottom) to the bath with $\beta=50$. 
}
\label{neps_t}
\end{figure}

The shape of the peak in $A(\omega, t)$ near $\omega=0$ indicates that  the system does not approach
the heavy Fermi liquid through a sequence of equilibrium states with decreasing temperature. Instead, 
it first establishes some kind of fairly stable non-thermal ``precursor'' state, which slowly 
evolves into the heavy Fermi liquid. In fact, comparison of the time-dependent spectra to equilibrium 
spectra at higher temperatures (Fig.~\ref{spectra_j1.5}, right panels) shows that for increased
temperature the quasi-particle peak would be broadened (and thus its maximum is shifted 
towards higher frequency), while in the time-dependent spectra, the location of the peak 
stays almost constant for large times. The larger amplitude of the peak in the precursor state can also not be explained 
with the small, transient change in doping - this would rather suggest a less prominent quasi-particle 
peak, since for large times the doping is slightly smaller than in the final state (Fig.~\ref{timeev_j1.5}). 
To further illustrate the difference between the non-thermal state with enhanced quasi-particle peak and the true FL equilibrium state, 
we plot in Fig.~\ref{neps_t}  the time-evolution of the momentum distribution function 
$n(\epsilon_k,t)$. 
The application of the pulse destroys the step-like feature marking the 
large Fermi surface. 
The fast relaxation of the singlet occupation and density back to the thermal values is associated with a partial recovery of the step 
in $n(\epsilon_k,t)$, but it is clear from Fig.~\ref{neps_t} that after $t=20$ ($\gamma=0.6$) or $t=50$ ($\gamma=0.2$) the distribution 
function has not yet thermalized. The thermal distribution function with its well-defined step feature is only recovered once the heavy 
quasi-particle band has been fully reconstructed.

\section{Summary and Conclusion}
\label{outlooksection}

We studied the relaxation dynamics of the Kondo lattice model (restricted to paramagnetic phases) using the nonequilibrium dynamical mean field formalism and an NCA impurity solver which exactly treats an eight-dimensional local problem consisting of a spin and its associated conduction electron orbital. 
This approach is well-suited to describe the formation of local singlets, which are 
favored 
by antiferromagnetic $J$.  
Comparison to data obtained with the numerically exact CT-HYB solver showed that the NCA approximation yields qualitatively correct results in equilibrium over a wide doping and temperature regime. In particular, it captures the crossover from a local moment regime into the heavy Fermi liquid regime as temperature is lowered in the doped system, with two crossover scales $T_K$ (opening of a pseudo-gap in the conduction electron spectral function) and $T^*$ (shift in the real part of the conduction electron self-energy and formation of a large Fermi surface).   

We determined the relaxation times in these various phases and crossover regimes by applying a weak magnetic field pulse which perturbs the local singlets. 
While these numbers are related to equilibrium properties of the system, they may still be nontrivial to extract from a conventional 
imaginary-time equilibrium calculation. 
In the doped system, the relaxation time grows with decreasing temperature, approximately proportional to the real part of the conduction electron self-energy. The slow relaxation in the heavy Fermi liquid phase is thus clearly associated with the existence of coherent heavy quasi-particles, and not, for example, with the presence of a pseudo-gap in the conduction electron spectral function. The relaxation time was also found to depend strongly on doping, with long relaxation times in the weakly doped heavy Fermi liquid regime. In contrast to the Hubbard model, the relaxation times in the (Kondo) insulating state are substantially shorter than in the weakly doped regime, at least for $J$ comparable to the hopping. 

To study the relaxation dynamics of strongly excited systems we considered quenches of the Kondo coupling and strong magnetic field pulses. Such perturbations lead to a considerable heating and it is thus easy to simulate the destruction of the low-temperature heavy-Fermi liquid state. By computing the time-dependent momentum distribution function for quenches from intermediate to small $J$, we could demonstrate the destruction of the step feature associated with the large Fermi surface within a time of a few inverse hopping and the shift from a large to a small Fermi surface on a slower time scale (the relaxation time of the thermal state). 

In order to demonstrate the formation of the large Fermi surface upon cooling, we simulated the time-evolution after a strong magnetic field pulse in the presence of a thermal particle reservoir, which removed the excess energy injected by the pulse. While a strong coupling to the heat-bath leads to a faster relaxation, it also smears out the heavy quasi-particle band. Nevertheless, for large enough Kondo coupling and large enough doping, one can have a well-defined large Fermi surface at low temperatures and fast relaxation. By computing the time-dependent momentum distribution functions and spectral functions we could show that the relaxation back to the thermal state happens in two stages: after a fast initial relaxation (whose time-scale depends on the strength of the coupling to the bath), a precursor state to the heavy Fermi liquid is formed, as evidenced by the appearance of a peak near the Fermi energy in the conduction electron spectral function, and a partially reconstructed step feature in the momentum distribution function. This fast relaxation is followed by a slower dynamics (presumably on time scales controlled by the long relaxation time in the heavy Fermi liquid), which leads to the thermalization of the narrow quasi-particle band also in the close vicinity of the Fermi energy.

The existence and the nature of the precursor state should certainly be corroborated and further studied in future investigations. 
For example, it would be desirable to systematically look at those states at 
a slightly weaker coupling to the environment in order to 
reduce the influence of the bath on the FL properties, in particular the quasi-particle weight and FL relaxation rate. 
However, smaller dissipation requires larger simulation times, which are not accessible within our current (single-processor)
implementation of the NCA equations. 
Furthermore, it would be desirable, although computationally quite expensive, to check the influence of the NCA approximation 
through a comparison to real-time results from higher order implementations of the self-consistent strong-coupling formalism. 

The calculations in this paper were limited to the paramagnetic phases of the antiferromagnetic Kondo lattice model. Many heavy electron materials however are close to a magnetic instability. It would be interesting to extend our study to symmetry broken phases in order to enable an interplay between electronic and magnetic excitations.

\acknowledgements
We thank 
J.~Otsuki and Th.~Pruschke
for useful discussions. The calculations were run on the Brutus cluster at ETH Zurich. We acknowledge support from the Swiss National Science Foundation (Grant PP0022-118866) and FP7/ERC starting grant No. 278023.

\begin{appendix}
\section{DMFT self-consistency with bath}

In this Appendix we briefly explain how a thermal fermionic
bath can be incorporated into the DMFT  equations for a semielliptic 
density of states.

The local $c$-electron Green function of the impurity model, 
$G_c(t,t')=-i \text{Tr} [T_\mathcal{C} e^{\mathcal{S} }c(t)c^\dagger(t')]/ \text{Tr} [T_\mathcal{C} e^{\mathcal{S} }]$
implicitly defines the self-energy $\Sigma$ via the impurity Dyson equation
\begin{equation}
G_c(t,t') = [i\partial_t + \mu - \Delta(t,t')-\Sigma(t,t')]^{-1}.
\label{dysimp}
\end{equation}
Here and in the following, time arguments are on the Keldysh contour $\mathcal{C}$, and $T_\mathcal{C}$ is the 
time-ordering operator on $\mathcal{C}$. We use the notation for Keldysh equations detailed in 
Ref.~\onlinecite{Eckstein10quench}. Momentum-dependent lattice Green functions 
$G_{\boldsymbol{k}}(t,t')=-i\langle T_\mathcal{C} c_{\boldsymbol{k}} (t)c_{\boldsymbol{k}}^\dagger(t') \rangle$ 
are then obtained from the lattice Dyson equation
\begin{equation}
G_{\boldsymbol{k}} (t,t') = [i\partial_t + \mu - \epsilon_{\boldsymbol{k}} -\Sigma(t,t')]^{-1},
\label{lattdys}
\end{equation}
where $\epsilon_{\boldsymbol{k}}$ is the noninteracting dispersion. 
From these functions the time-dependent momentum distribution can be obtained,
\begin{equation}
n(\epsilon_{\boldsymbol{k}},t) = -i G_{\boldsymbol{k}}^< (t,t).
\label{nk_definition}
\end{equation}
For a lattice whose noninteracting density of 
states is semi-elliptical with bandwidth $4v$, the local (momentum averaged) lattice Greenfunction can be shown 
to satisfy the self-consistent equation
\begin{align}
G(t,t') &\equiv \sum_{\boldsymbol{k}} G_{\boldsymbol{k}} (t,t') 
\\
&= 
[i\partial_t + \mu - v^2 G(t,t')   -\Sigma(t,t')]^{-1}.
\end{align}
Note that the second equality does not depend on the form of the self-energy, but only 
on the distribution of the band-energies $\epsilon_{\boldsymbol{k}}$.\cite{SFB}
Comparison with the impurity Dyson equation (\ref{dysimp}) then yields the standard
DMFT self-consistency condition, Eq.~(\ref{self-consistency}).

If an additional fermionic reservoir is coupled to the lattice at every site, one has to modify 
the lattice Dyson equation (\ref{lattdys}) by adding the hybridization function $\Delta_\beta(t,t')$
to the free dispersion,
\begin{equation}
G_{\boldsymbol{k}} (t,t') = [i\partial_t + \mu - \epsilon_{\boldsymbol{k}}-\Delta_\beta(t,t') -\Sigma(t,t')]^{-1}.
\label{lattdys2}
\end{equation}
Hence the closed form equation for the momentum averaged Green function becomes
\begin{equation}
G(t,t') = 
[i\partial_t + \mu - v^2 G(t,t') -\Delta_\beta(t,t')  -\Sigma(t,t')]^{-1},
\end{equation}
and, by comparison with Eq.~(\ref{dysimp}), we obtain the DMFT self-consistency with fermionic bath,
\begin{equation}
\Delta(t,t') = v^2 G(t,t') + \Delta_\beta(t,t').
\end{equation}

\end{appendix}


\begin{thebibliography}{99}
\bibitem{footnote00}
If a smaller local Hilbert space is used, this physics requires
the summation of higher-order diagrams beyond the NCA.
\bibitem{footnote01}
To avoid oscillations due to a cutoff in the Fourier integral in Eq.~(\ref{A_omega_t}), we have extrapolated $G^\text{ret}(t',t)$ 
to large times using an exponential function. This is possible, because in the model with bath, the oscillations in $G^\text{ret}(t',t)$ 
are damped rather quickly.
\bibitem{footnote02} In the presence of the bath the FL relaxation rate is somewhat reduced.
\end{thebibliography}
\end{document}